\documentclass[twocolumn,amsmath,amssymb,nofootinbib]{revtex4}
\usepackage{graphicx}% Include figure files
\usepackage[latin1]{inputenc}
\usepackage{dcolumn}% Align table columns on decimal point
\usepackage{bm}% bold math
\usepackage{epsfig}
\usepackage{xcolor}
\usepackage{amsmath}
\usepackage{algorithm}
\usepackage[noend]{algpseudocode}

\makeatletter
\def\BState{\State\hskip-\ALG@thistlm}
\makeatother

\begin{document}

\preprint{Draft -- version 5}

\title{Visibility graphs for image processing}% Force line breaks with \\

\author{Jacopo Iacovacci$^{1,2}$ and Lucas Lacasa$^3$}
%\email{email}
\affiliation{$^{1}$Department of Surgery and Cancer, Division of Computational and Systems Medicine, 
Imperial College London, London SW72AZ (United Kingdom)}
\affiliation{$^{2}$The Molecular Biology of Metabolism Laboratory, The Francis Crick Institute, London NW11AT (United Kingdom)}%
\affiliation{$^{3}$School of Mathematical Sciences, Queen Mary University of London, E14NS London (United Kingdom)}%


\begin{abstract}
The family of image visibility graphs (IVGs) have been recently introduced as simple algorithms by which scalar fields can be mapped into graphs. Here we explore the usefulness of such operator in the scenario of image processing and image classification. We demonstrate that the link architecture of the image visibility graphs encapsulates relevant information on the structure of the images and we explore their potential as image filters and compressors. We introduce several graph features, including the novel concept of Visibility Patches, and show through several examples that these features are highly informative, computationally efficient and universally applicable for {general pattern recognition and image classification tasks}.

%(i) image compression, (ii) feature extraction for pattern recognition tasks and (iii) .

\end{abstract}

\pacs{}% PACS, the Physics and Astronomy
\keywords{} \maketitle

\section{Introduction}
%\textcolor{blue}{Some general sentences about image processing and transformation from data to graphs.}\\
Visibility and Horizontal Visibility graphs (VG/HVGs) were proposed in \cite{PNAS, PRE, multivariate} as a family of simple mappings between ordered sequences (e.g. time series) and graphs \cite{vito}, this being one possible strategy to perform graph-theoretical time series analysis (see \cite{Kurths2017} and references therein for a recent overview). Consider an ordered sequence $\{{\bf x}(t)\}_{t=1}^N$, where ${\bf x}(t)\in \mathbb{R}^m$. For $m=1$ a typical case of such sequence are univariate time series describing the activity of some system or one-dimensional dynamical systems, whereas for $m>1$ we consider multivariate time series or trajectories from high-dimensional dynamical systems. In the former case, a time series of $N$ data is mapped into a graph of $N$ nodes such that two nodes are linked in the graph if particular geometric and ordering criteria hold in the sequence (in the multivariate case, there are $m$ replicas of the set of nodes --each of these replicas representing a different layer-- and the edge set is a priori different for each layer, thereby defining a {\it multiplex visibility graph} \cite{multivariate, multiplex1, multiplex2}). This mapping enables the possibility of performing graph-theoretical time series analysis and hence permits to build bridges between the theories of dynamical systems, signal processing and graph theory.\\ 

\noindent Theoretical research on visibility graphs has elaborated on mathematical methods \cite{severini, Luque2016,nonlinearity,motifs} and some rigorous results on the properties of these graphs when associated to canonical models of complex dynamics have been obtained \cite{epl, jns, pre2013,quasi}. From a practical point of view, this method has been used as a feature extraction procedure to construct feature vectors from time series for statistical learning purposes (see \cite{Shao2010,Ahmadlou2010,Ahmadlou2012,Bhaduri2016,Sanino,meditation_VG,Uri} for just a few examples in the life sciences or \cite{physics3,fluiddyn0,fluiddyn1,fluiddyn2,physics2,suyal,Zou,physio1} for other applications in the physical sciences).\\
\noindent Very recently \cite{scalar}, this paradigm has been theoretically extended to handle scalar fields. Here we consider the particular two-dimensional case $h(x,y):\mathbb{R}^2\to \mathbb{R}$ (this being conceptually closer to the original context of visibility graphs analysis as defined in Urban Planning \cite{VG}) and explore from a theoretical and practical point of view its usefulness in image processing and pattern recognition, {proposing in this way a novel methodology in the field of {\it graph-based image analysis} \cite{imageprocessgraphs}}. We will show that, under this extension, visibility graphs can be extracted from images and subsequently used both for image filtering and compression, and as a universal {feature extraction} protocol for supervised learning purposes, complementing more standard approaches in the field \cite{rafa}.\\
The rest of the paper goes as follows: in section II we define the method of image visibility graphs and present some of its properties, along with a few key metrics which will be relevant as graph-theoretical image descriptors. In section III we explore the performance of image visibility graphs in the tasks of image filtering and compression, and in section IV we evaluate their performance in the tasks of pattern recognition and texture classification. In section V we conclude.

%In the former case, an image is projected into a feature space, such that classification tasks and pattern recognition tasks (face detection, edge detection, human detection) can be addressed using these novel set of features. In this case, we need to evaluate (i) how informative the new set of features is, as compared to traditional ones, and (ii) how orthogonal the new set of features is, compared to standard benchmark. This last question can be assessed via PCA/ICA. If on the other hand we consider this mapping as an image filter, then we need to assess whether standard image processing tasks are display better performance on the filtered images than on the original ones, and compare such improvement with respect to state of the art filters.\\
%Before addressing these applications, we shall introduce the algorithms along with a few basic properties.\\

\section{Image Visibility graphs}
For completeness, we start by introducing the definitions of two types of visibility graphs extracted from univariate time series, and one of their possible extensions to images.\\

\noindent {\bf Definition (VG). } {\it Let ${\cal S}=\{x_1,\dots,x_N\}$ be a ordered sequence of $N$ real-valued, scalar data. A Visibility Graph} (VG) {\it is an undirected graph of $N$ nodes, where each node $i$ is labelled according to the time order of its corresponding datum $x_i$. Hence $x_1$ is mapped into node $i=1$, $x_2$ into node $i=2$, and so on.
Then, two nodes $i$ and $j$ (assume $i<j$ without loss of generality) are connected by an (undirected) link if and only if one can draw a straight line connecting $x_i$ and $x_j$ that does not intersect any intermediate datum $x_k, \ i<k<j$. Equivalently, $i$ and $j$ are connected if the following {\it convexity} criterion is fulfilled:}
$$x_k< x_i + \frac{k-i}{j-i}[x_j-x_i],\ \forall k: i<k<j$$
A similar definition follows for a Horizontal Visibility Graph (HVG), but in this latter graph two nodes $i$, $j$ (assume $i<j$ without loss of generality) are connected by a link if and only if one can draw a {\it horizontal} line connecting $x_i$ and $x_j$ that does not intersect any intermediate datum $x_k, \ i<k<j$. Equivalently, $i$ and $j$ in the HVG are connected if the following {\it ordering} criterion is fulfilled:
$$x_k<\inf(x_i,x_j),\ \forall k: i<k<j.$$
One can now extend the definition to handle not just ordered sequences but images \cite{scalar}:\\

\noindent {\bf Definition (IVG). }{\it Let ${\cal I}$ be a $N\times N$ matrix where ${\cal I}_{ij}\in \mathbb{R}$. The {\it Image Visibility Graph}} (IVG) {\it is a graph of $N^2$ nodes, where each node is labelled by the indices of its corresponding datum ${\cal I}_{ij}$, such that two nodes $ij$ and $i'j'$ are linked if
\begin{itemize}
\item $i=i'$ $\text{OR}$ $j=j'$ \text{OR} $[i=i'+p\  \text{AND}\  j=j'+p]$, for some integer $p$, and
\item ${\cal I}_{ij}$ and ${\cal I}_{i'j'}$ are linked in the VG defined over the ordered sequence which includes $ij$ and $i'j'$.
\end{itemize}}
The {\it Image Horizontal Visibility Graph} (IHVG) follows equivalently if in the second condition we make use of HVG instead of VG. Note that the preceding definition indeed coincides with the definition of an IVG/IHVG in the so-called FCC extension class \cite{scalar}, this being one of other possibilities for the analysis of high-dimensional scalar fields, which we adopt here for convenience as this particular one is well-fitted for image processing.\\

\noindent In the rest of this section we will consider a few properties of IVG/IHVGs which will be of particular importance for the tasks of image filtering, compression and classification.
\begin{figure}[h!]
\includegraphics[width=1\columnwidth]{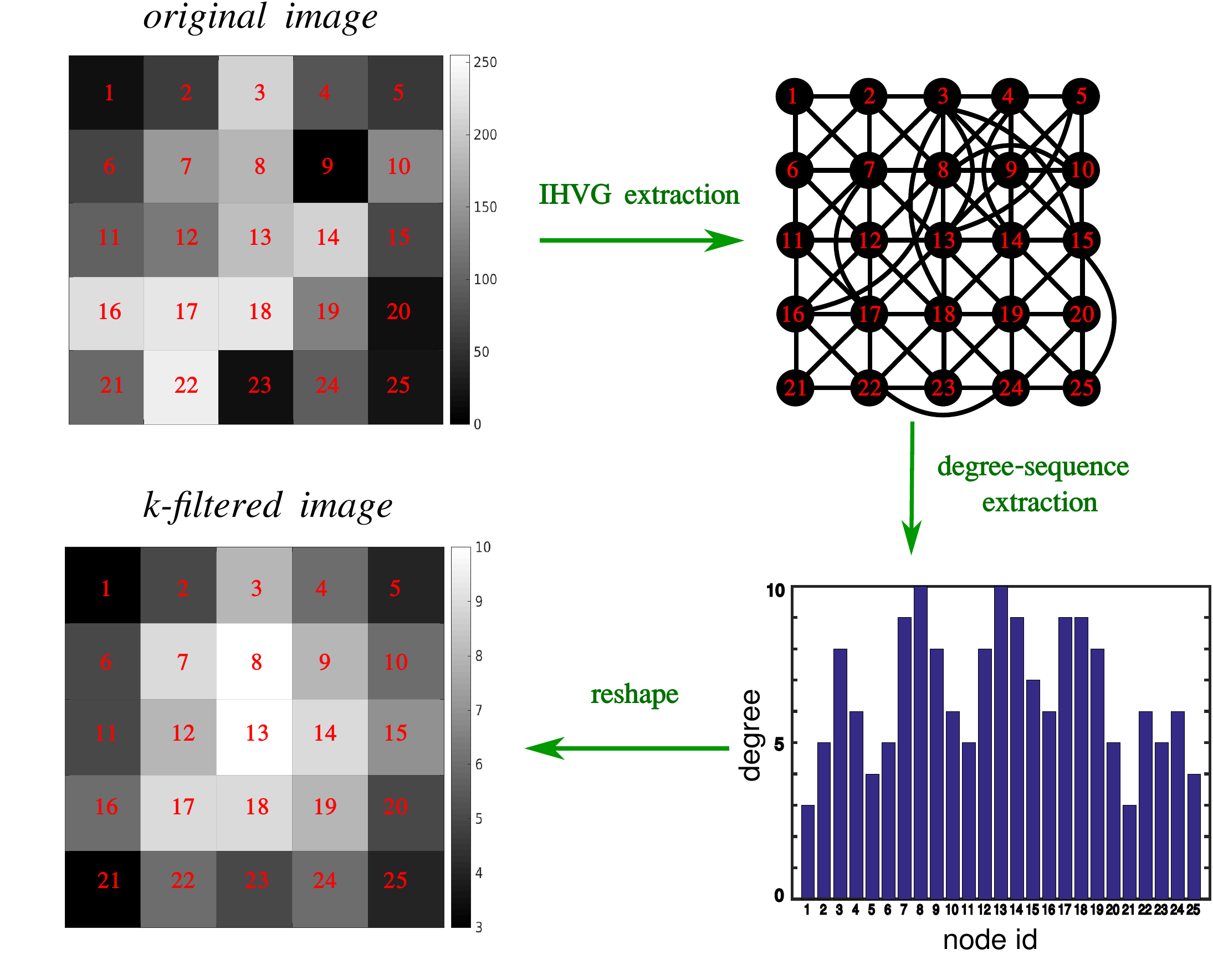}
\caption{{Illustration of the process of extracting a topological plot from an image. The image is transformed into its corresponding IVG/IHVG, where the nodes properties inherit the spatial local information of the pixels. For each node a specific topological property is measured --for example its degree-- and the measured values are mapped back to form a matrix of the same size of the original image.}}
\label{fig:plots}
\end{figure}

\subsection{Topological plots}
One of the simplest graph properties \cite{vito} is those related to local properties of nodes. For instance, the degree $k_i$ of node $i$ in an undirected graph is defined as the number of links incident to $i$, while the local clustering coefficient $c_i$ of node $i$ is defined as the percentage of nodes which are adjacent to $i$ which are also adjacent to each other. One can assign labels to the nodes of a graph and can accordingly define the degree sequence, clustering sequence, etc of a given graph.\\
In the realm of VG/HVGs nodes have a pre-specified (temporal) ordering, and therefore these sequences inherit such order: the degree sequence $(k_1,k_2,\dots)$ is such that $k_i$ is the degree of the node associated to datum $x_i$. These considerations can be extended to the image setting considered here, as similar labels can be attached to the nodes of an IVG/IHVG. By construction however, it is natural to order the node set of an IVG/IHVG in terms of an array (matrix) which represents the spatial positions of the associated pixels in the image. In this sense, one can therefore define the {\it degree plot}, the {\it node clustering plot}, etc, as primary topological properties of the IVG/IHVGs. {For illustration, in Figure \ref{fig:plots} we depict the process of extracting the degree plot of the IHVG associated to a $5\times5$ grayscale image. In the degree plot (also called {\it k-filtered image}, see section III) the pixel values $\{k_{ij}\}$ correspond to the degree of the nodes $ij$ in the IHVG.} \\
In section III we will use the topological plots for image filtering and compression. Before that, we need to introduce a few other features which will be relevant for image classification.

\subsection{Global features: Degree distribution}

Assuming the premise that the IVG/IHVG inherits information of the associated image, a natural procedure is to extract features from these graphs and use them for image classification purposes. Now, graph properties can be classified into global properties (accounting for topological information of the whole graph), or local properties (which account for properties of small substructures). Global features include the degree distribution $P(k)$, defined as the percentage of nodes that have degree $k$, and other properties of the degree sequence, clustering distribution, spectral properties, distribution of cycles, etc.\\
In the context of visibility graphs, a recent theorem \cite{Luque2016} proved that under suitable conditions, HVGs are in bijection with their degree sequence, suggesting that the degree distribution is indeed a relevant and informative global property. Additionally, it has been shown that $P(k)$ can be computed exactly for simple images and is a good feature to distinguish different sorts of spatio-temporal dynamical systems \cite{scalar}. Justified by these previous findings and for the sake of parsimony we will focus hereafter on $P(k)$ as the global feature under study.

\begin{figure}[h!]
\includegraphics[width=0.55\columnwidth]{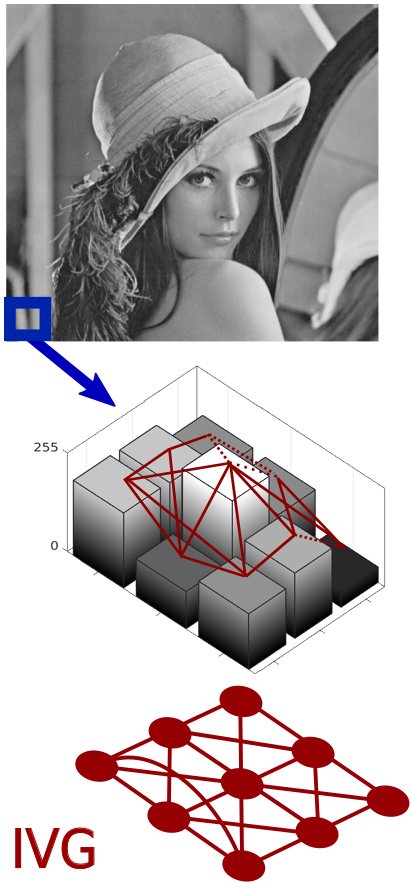}
\caption{Illustration of the process of measuring a visibility patch of order $p=3$.}
\label{fig:patch}
\end{figure}

\subsection{Local features: Visibility Patches}
Opposed to global features, we might be interested in studying local properties of the image visibility graphs, and in this subsection we introduce a set of novel features which aims at detecting such properties. The procedure consists in counting repetitions of small subgraphs in the IVG/IHVG associated to a given image (see Figure \ref{fig:patch} for an illustration). We call these subgraphs {\it Visibility Patches} (VPs) and we formally define them in the following way:\\ 
%which constitute a natural extension of the concept of sequential visibility motifs \cite{motifs, motifs2} to visibility graphs of spatially extended data. For the sake of formality, we now introduce the definition:\\

\noindent {\bf Definition (Visibility Patch)} {\it Consider an }\text{IVG} {\it of $N^2$ nodes with adjacency matrix ${\bf A}=\{A_{ij}\}_{i,j=1}^{N^2}$ associated to an $N\times N$ real-valued image matrix. A Visibility Patch of order $p$ }VP$_p$ {\it is any subgraph of the IVG formed by a set of $p^2$ nodes $\{ij\}_{i=s,j=s'}^{s+p-1,s'+p-1}$ for arbitrary $s,s'$ that satisfy $1\leq s,s' \leq N -p$.}\\

\noindent Note that the definition above extends naturally to {\it Horizontal} Visibility Patches of order p (HVP$_p$) if the graph under study is IHVG instead of IVG.
{The lowest order that yields nontrivial visibility patches is $p=3$. These visibility patches can be detected by sliding a cell of size $3\times3$ pixels with stride 1 along the entire image, and extracting the corresponding IVG/IHVGs within the cell, as shown in Figure \ref{fig:patch}. For the sake of exposition we will focus on $p=3$ in this work, but note that algorithms are scalable to higher orders if higher performance is needed.\\}
%Once we decompose an image in patches of equal size, we can easily extract the corresponding visibility patches, whose ordered sequence and whose topology will carry structural information on the image respectively on the global and on the local scale at a resolution scale dependent on the patch size. 
A relevant quantity that summarises the abundance of visibility patches is given by the frequency distribution of each of these patches, labelled {\it (Horizontal) Visibility Patch Profile}  {\bf Z} (we label this profile with the bold letter {\bf Z}  indistinctively for patches computed on IHVG and IVG, and context will suggest which one of the two it applied to).\\

\noindent Interestingly, note that the concept of visibility patch is the natural extension to images of the concept of sequential visibility graph motifs \cite{motifs, motifs2}. As a matter of fact, algorithmic computation of visibility patches reduces to checking for the presence or absence of certain combination of visibility graph motifs. This fact enables mathematical tractability of Visibility Patches as well as the design of a linear time algorithm to extract visibility patches, as we show below.\\

\begin{figure}[h!]
\includegraphics[width=0.75\columnwidth]{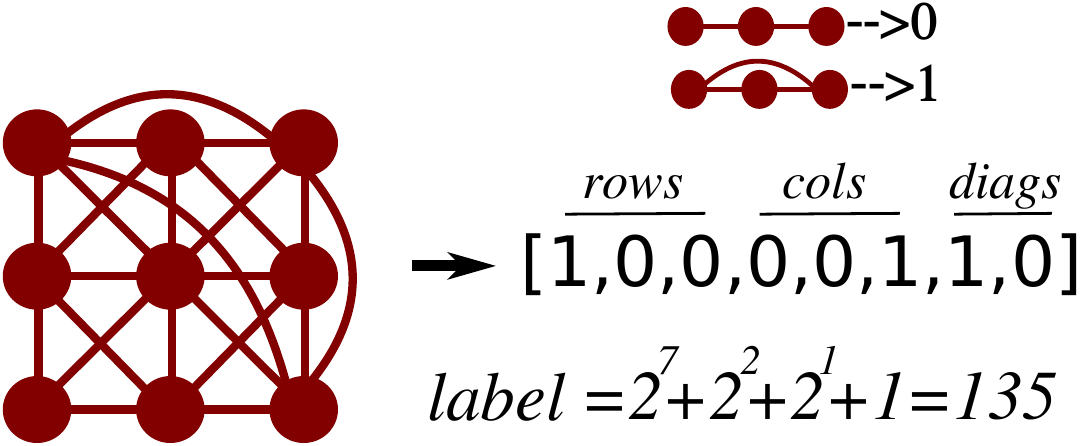}
\caption{Illustration of the enumeration of a concrete HVP$_3$ in terms of the sequential visibility graph motifs of order 3 (see table \ref{HVG_motifs}).}
\label{fig:patch3x3}
\end{figure}

\begin{table*}
%\begin{ruledtabular}
\begin{tabular}{|cc|c|c|}
\hline
{\bf Label}&{\bf Motif}&{\bf HVG}&{\bf VG}\\
\hline
0&\begin{minipage}{.1\textwidth}
\includegraphics[width= 0.5\textwidth]{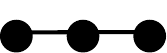}
\end{minipage}
&$\{ \forall x_0,x_2: x_1>x_0 \}\cup\{ \forall x_0: x_1<x_0,x_2<x_1\} $& $\{ \forall x_0, x_1:x_2\leq 2x_1-x_0\}$ \\
1&\begin{minipage}{.1\textwidth}
\includegraphics[width= 0.5\textwidth]{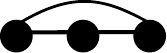}
\end{minipage}& $\{\forall x_0: x_1<x_0, x_2>x_1\}$&$\{ \forall x_0, x_1:x_2> 2x_1-x_0\}$ \\
\hline
%1&\begin{minipage}{.1\textwidth}
%\includegraphics[width= 0.7\textwidth]{4motif-1.pdf}
%\end{minipage}&$\{ \forall (x_0,x_1), x_2<x_1, x_3<x_2\}\cup\{\forall (x_0,x_3), x_1>x_0,x_2>x_1\} $\\
%2&\begin{minipage}{.1\textwidth}
%\includegraphics[width= 0.7\textwidth]{4motif-2.pdf}
%\end{minipage}& $\{\forall x_0, x_1<x_0,x_2=x_1,x_3>x_2\}$ \\
%3&\begin{minipage}{.1\textwidth}
%\includegraphics[width= 0.7\textwidth]{4motif-3.pdf}
%\end{minipage}& $\{\forall x_0, x_1<x_0, x_1<x_2<x_0,x_3<x_2\}\cup\{\forall (x_0,x_3),x_1<x_0,x_2>x_0\}$\\
%4&\begin{minipage}{.1\textwidth}
%\includegraphics[width= 0.7\textwidth]{4motif-4.pdf}
%\end{minipage}& $\{ \forall x_0, x_1>x_0,x_2<x_1, x_3>x_2  \}\cup\{\forall x_0, x_1<x_0, x_2<x_1, x_2<x_3<x_1\}$\\
%5&\begin{minipage}{.1\textwidth}
%\includegraphics[width= 0.7\textwidth]{4motif-5.pdf}
%\end{minipage}&$\{\forall x_0, x_1<x_0, x_1<x_2<x_0, x_3>x_2\}$\\
%6&\begin{minipage}{.1\textwidth}
%\includegraphics[width= 0.7\textwidth]{4motif-6.pdf}
%\end{minipage}& $\{\forall x_0, x_1<x_0, x_2<x_1, x_3>x_1\}$\\
%\hline
\end{tabular}
\caption{Sequential VG/HVG motifs of order $3$. Each motif can be characterized according to a set of inequalities in the associated time series (see \cite{motifs, motifs2} for details on extension to higher orders).}
	\label{HVG_motifs}
\end{table*}

\noindent {\bf Enumeration. } By construction, an (Horizontal) Visibility Patch of order $p$, (H)VP$_p$, is composed by an ordered sequence of $2p+2$ sequential visibility graph motifs of order $p$ \cite{motifs, motifs2}, along the $p$ rows, $p$ columns and the main diagonal and anti-diagonal of the patch, supplemented by an additional number $q=4(p-2)$ of motifs of lower order associated to the off-diagonals and off-anti-diagonals: 4 motifs of order $p-1$, 4 motifs of order $p-2$, and so on. Accordingly, a natural enumeration of a given (H)VP$_p$ is via a (2p+2+q)-dimensional vector $\bf V$ given by
$${\bf V}=[r_1,r_2,\dots,r_p,c_1,c_2\dots,c_p,d_1,d_2,\dots],$$where each element of this vector enumerates the label of the actual visibility graph motif in the direction it represents.
For example, in figure \ref{fig:patch3x3} we consider an example for the simpler nontrivial case $p=3$. There are only two possible (horizontal) sequential visibility graph motifs of order 3 (see table \ref{HVG_motifs}) --which we can label 0 and 1-- and a single possibility for motifs of order two --and therefore these do not contribute to the combinatorics--. This yields a total of $2^{(2p+2)}=256$ different possible patches, which can be enumerated by a $2p+2=8$ dimensional vector. As there are just two non-trivial motifs of order 3, one can enumerate all possible patches of order $3$ by interpreting $\bf V$ as the binary expansion of its label, i.e.
$${{\cal L}[{\bf V}]}=r_12^7 + r_22^6 + \dots + d_1 2^1 + d_2 2^0 +1,$$
such that for the particular case reported in Fig. \ref{fig:patch3x3}, we find the motif label 0 in all rows, columns and diagonals except the first row, the last column and the main diagonal, i.e. ${{\cal L}[{\bf V}]}=2^7+2^2+2^1+1=135$, i.e. corresponding to patch no. 135.\\

\begin{figure}[h!]
\includegraphics[width=0.9\columnwidth]{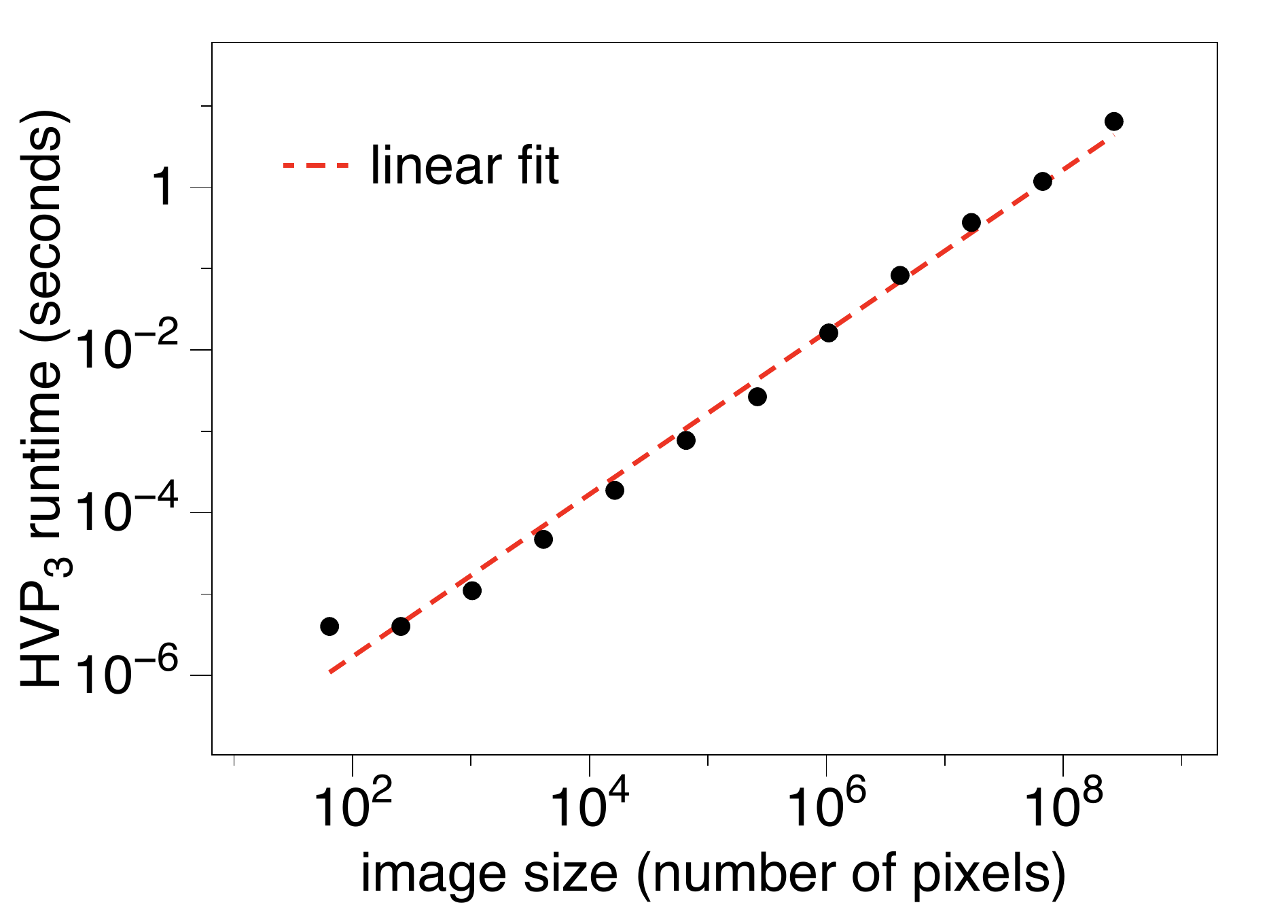}
\caption{Running time of the Horizontal Visibility Patch Profile algorithm of order 3 as a function of the size of the image. Feature extraction scales linearly (computation performed on a 2.5GHz IntelCore i7 processor).}
\label{fig:runtime}
\end{figure}

\noindent {\bf Linear time recognition algorithm. } By further capitalizing on the relation between Visibility Patches and Visibility motifs, we can build an algorithm to estimate the Visibility Patch profile {\bf Z} of a given image which scales linearly with the number of pixels of the image. In particular, recognition of motifs of any order $p$ can be done in linear time by using the inequality set in Table \ref{HVG_motifs} (see \cite{motifs,motifs2} for an explicit inequality set for $p=4$). For an illustration, consider the concrete case of HVP$_3$. We initially define ${\bf V=0}$ for each patch, and then update the 1s in each patch according to the Algorithm 1 displayed beside, updating the HVP$_3$ profile $\bf Z$ for each pair $i,j$ (computation of VP$_3$ would follow equivalently).\\ 
In Fig. \ref{fig:runtime} we plot how runtime scales as a function of the size of the image (in number of pixels), for randomly-generated grayscale images $X_{ij}\sim {\cal U}\{0,255\}$, showcasing a linear time complexity and therefore an efficient feature extraction.\\ 

\begin{algorithm}
\caption{-- HVP$_3$ profile $\bf Z$}\label{euclid}
\begin{algorithmic}[1]
\Procedure{HVP$_3$}{}
        
\State $X_{i,j} \gets image$

\BState \emph{Rows}:
\If {$X_{i,j+1}<H_{i,j} \ \text{\bf{and}\ } X_{i,j+2}>X_{i,j+1}$} $row \gets 1$
\EndIf

\BState \emph{Columns}:
\If {$X_{i+1,j}<H_{i,j} \ \text{\bf{and}\ } X_{i+2,j}>X_{i+1,j}$} $col \gets 1$
\EndIf

\BState \emph{Diagonal}:
\If {$X_{i+1,j+1}<H_{i,j} \ \text{\bf{and}\ } X_{i+2,j+2}>X_{i+1,j+1}$} $diag \gets 1$
\EndIf

\BState \emph{Anti-Diagonal}:
\If {$X_{i+1,j+1}<H_{i,j+2} \ \text{\bf{and}\ } X_{i+2,j}>X_{i+1,j+1}$} $adiag \gets 1$
\EndIf

\\
        
\noindent {\bf for} $k=1:N-3:3$
\State $i \gets {k}$
\State $j \gets {k}$
\State{\bf call} {\it Rows}
\State{\bf call} {\it Columns}
\State$i\gets i+1$ {\bf Repeat} {\it Rows}
\State$i\gets i+1$ {\bf Repeat} {\it Rows}
\State $i \gets {k}$
\State$j\gets i+1$ {\bf Repeat} {\it Columns}
\State$j\gets i+1$ {\bf Repeat} {\it Columns}
\State $j \gets {k}$
\State{\bf call} {\it Diagonal}
\State{\bf call} {\it Anti-Diagonal}
\State {\bf update} Profile $\bf Z$
\EndProcedure
\end{algorithmic}
\end{algorithm}

\noindent Theoretical analysis of visibility patches, including analytical results for the visibility patch profiles, is in principle possible and will be published elsewhere \cite{inprep}. Here we are interested in exploring the performance of these features in practical situations. In section IV we will explore the performance of global features (e.g. degree distribution) and local features (patch profiles of order 3, as computed from (H)VP$_3$, these being the simplest nontrivial patches) in different image classification tasks.

%To extract an ideal $local$ feature vector from an image we can take the distribution of the visibility patches of size NxN nodes so that each feature corresponds to the probability of appearance of a given visibility patch and the number of features (vector length) equals the admissible number of patches of size NxN. \\  

%In the following we will always consider as a local descriptor the distribution of Visibility Patches of size 3x3, which correspond to the finest resolution of the image in terms of nontrivial visibility graph and and whose number is easy to handle computationally. 

\begin{figure*}[htpb!]
\includegraphics[width=1.6\columnwidth]{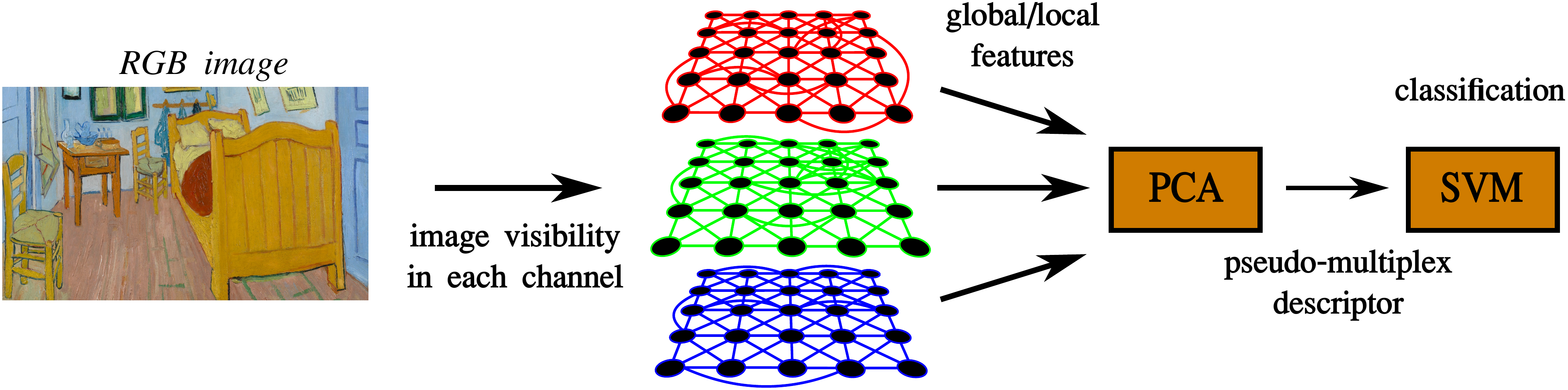}
\caption{Illustration of the process of extraction an image multiplex visibility graph (here applied to a RGB image, hence yielding a multiplex graph with $m=3$ layers). Features extracted from each of the layers are combined using a Principal Component Analysis to obtain a pseudo-multiplex descriptor, that is fed into a classifier.}
\label{fig:multiplex}
\end{figure*}

\subsection{Multiplex features}
To round off this section, we now consider the situation where one might wish to handle a set of images at once. Examples include (i) RGB images, where a given image is actually built by combining three channels (R, G and B), (ii) a given image filtered in different ways (e.g. a set of different topological plots, or image filters at different frequency bands, etc), (iii) a `temporal' image comprised of a certain number of snapshots, etc. In every case, the image under study is actually a set of $m$ images ${\cal I}=\{{\cal I}^{(1)},\dots,{\cal I}^{(m)}\}$ (for an RGB image, $m=3$). One might then be interested in extracted the image visibility graph of each sub-image separately, or construct the image visibility graph of the image set at once. In this latter case, this naturally leads to the construction of an {\it Image Multiplex Visibility Graph}, i.e. the image version of a multiplex visibility graph \cite{multivariate}:\\

\noindent {\bf Definition (Image Multiplex Visibility Graph)} {\it Let ${\cal I}=\{{\cal I}^{(1)},\dots,{\cal I}^{(m)}\}$ be an ordered sequence of $m$ $N\times N$ grayscale images. The image multiplex visibility graph is a multiplex visibility graph of $m$ layers, where layer $\alpha=1,\dots, m$ encodes the image visibility graph of grayscale image ${\cal I}^{(\alpha)}$.}\\

{\noindent Note that the resulting graph is a multiplex network, a specific type of multilayer network \cite{multiplex1, multiplex2} characterised by a unique set of nodes which is connected differently on several independent layers or networks (see \cite{multilayerCN} for another recent proposal for multilayer graph-based analysis of RGB images). This architecture is possible thanks to the natural alignment of every pixel in each sub-image (${\cal I}^{(\alpha)}_{ij}$ aligned with ${\cal I}^{(\alpha')}_{ij}\  \forall(\alpha,\alpha')$). %Then, intra-layer connections between nodes are determined by their corresponding pixel intensities in the specific sub-images.
As it is customary in multiplex networks \cite{multiplex1,multiplex2,battiston}, one can proceed to extract features from the image multiplex visibility graph of several kinds:
\begin{enumerate}
\item {\it Intra-layer descriptors}: these are any graph measures computed independently on each layer \cite{vito,boccaletti2014structure} (e.g. global or local features such as the ones discussed above).
\item {\it Inter-layer descriptors}: these are in general measures of correlation or interdependence of the features between different layers; examples include the edge overlap \cite{bianconi2013statistical}, degree-based correlation measures \cite{nicosia2015measuring,multivariate},  and correlations between network mesoscale descriptors such as node clusters \cite{iacovacci2015mesoscopic}.
\item {\it Intrinsically multiplex descriptors}: these are features that can be extracted only by considering the overall multiplex architecture of the network \cite{Iacovacci2016,cardillo2013emergence}, for example multiplex clustering coefficients and multiplex motifs \cite{battimotifs,cozzo}, node multiplex centrality measures \cite{fmPR, mPR}, and multiplex communities \cite{4minutes,de2015identifying}. 
\end{enumerate}
In this work we explore the performance of efficiently combining simple intra-layer descriptors to obtain {\it pseudo-multiplex descriptors}. Indeed, a natural procedure is to take into account the whole multiplex structure and process in parallel the structural information contained in each layer by applying a method of dimensionality reduction such as Principal Component Analysis to project the concatenated feature vector of intra-layer descriptors into an adequate subspace (see Fig. \ref{fig:multiplex} for an illustration). In section IV we will illustrate the use of image multiplex visibility graphs as a feature extraction method for texture classification.}

\section{Filtering and Compression}

\subsection{Robustness to noise}
\vspace{-2mm}
We initially explore the robustness against noise pollution of the family of image visibility graphs. We consider the standard grayscale Lena image and pollute it with white Gaussian noise of different power. We then measure, for each level of noise to signal ratio (NSR), the similarity between the non-polluted and the polluted case. For that sake we measure the overall link overlap ${\cal Q}(I,I^{\text{NSR}})$ between the image visibility graphs extracted from the original image and the respective image visibility graph extracted from the image contaminated with noise, defined as:

\begin{equation}
{\cal Q}({\cal I},{\cal I}^{\text{NSR}})=\frac{ \sum_{ij}A_{ij}A^{\text{NSR}}_{ij} - {\cal Q}_{\text{min}} }{ \max[\sum_{ij}A_{ij},\sum_{ij}A^{\text{NSR}}_{ij}]-{\cal Q}_{\text{min}}}
\end{equation}
where ${\bf A}=\{A_{ij}\}$ and ${\bf A}^{\text{NSR}}=\{A_{ij}^p\}$ denote the adjacency matrices associated to the image visibility graph from the original image and the polluted image, respectively.  ${\cal Q}_{\text{min}}$ is the number of links that by construction two IVG/IHVGs of the same size always share because of the common lattice structure: ${\cal Q}_{\text{min}}=2(N-1)N+2(N-1)^2$. The overlap ${\cal Q}$ could be used as a general indicator of the similarity between two images of the same size or of the quality of a given image after a certain noise or filter is applied, however it doesn't saturate to zero when comparing, say, two instances of white noise. We can build a well-defined similarity function as it follows:
\begin{equation}
{\cal S}_{\text{IVG/IHVG}}({\cal I},{\cal I}^{\text{NSR}})=\frac{ {\cal Q}({\cal I},{\cal I}^{\text{NSR}}) - {\cal Q}_\infty}{ 1-{\cal Q}_\infty},
\end{equation}
where ${\cal Q}_\infty:={\cal Q}({\cal I},{\cal I}^{\text{NSR}=100})$. By construction, this similarity measure is equal to one for equal images and reaches zero when the image is polluted with noise with a $\text{NSR}=100$.\\

\begin{figure}[h!]
\includegraphics[width=1\columnwidth]{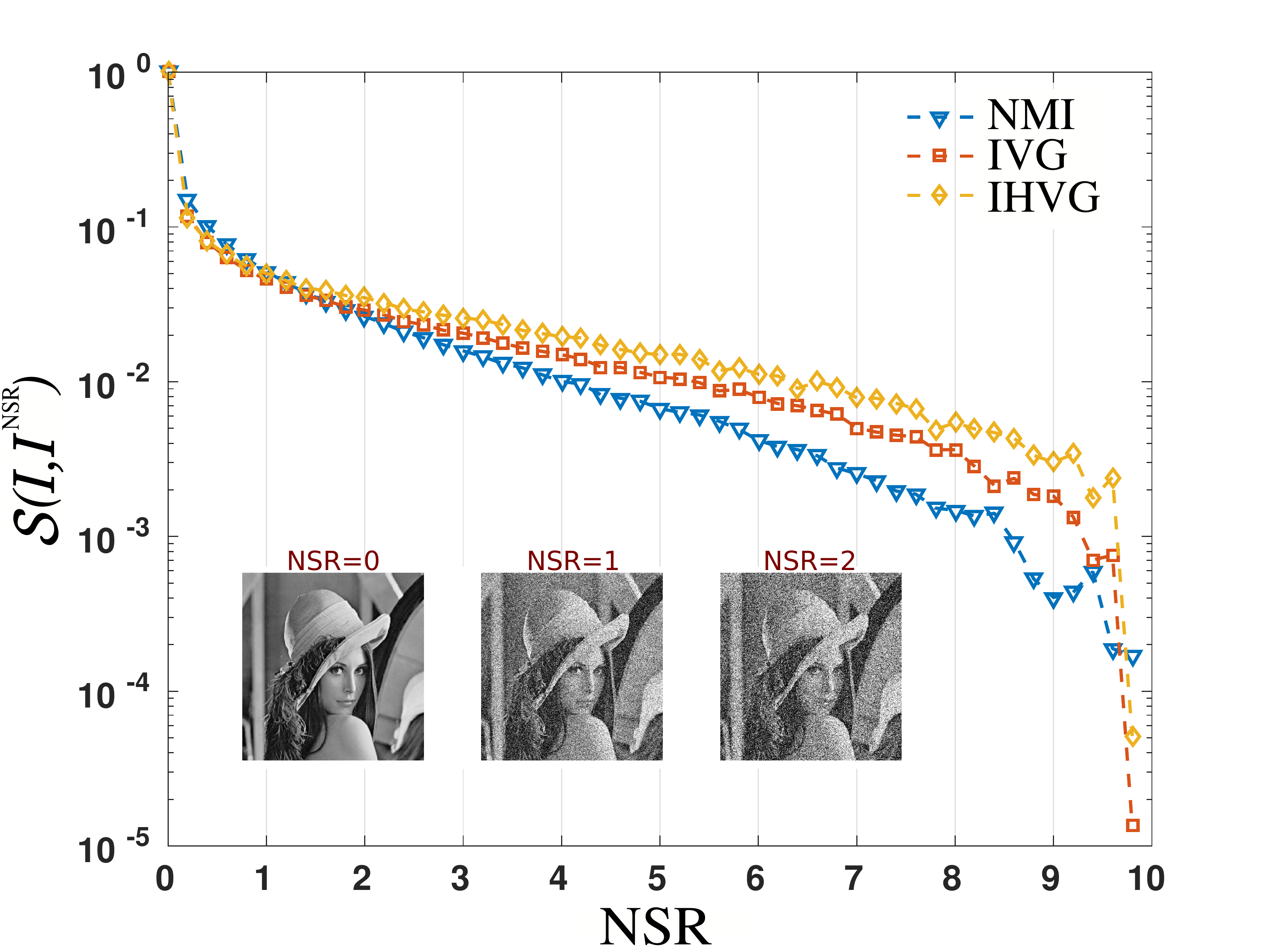}
\caption{Semi-log plot of Similarity measure between the IVG/IHVG of Lena and Lena polluted with a certain amount of Gaussian white noise, as a function of the Noise-to-Signal Ratio (NSR). The benchmark similarity measure computed directly on images is based on an appropriately rescaled version of the Normalized Mutual Information. Information mapped to the IVG/IHVG setting seems to be more robust to noise contamination than the raw information stored in the image.}
\label{fig:similarity}
\end{figure}

\noindent In Fig.\ref{fig:similarity} we plot in semi-log scale $\cal S$ (computed in both IVG and IHVG) as a function of the NSR. For comparison, we also examine the robustness against noise directly on the images by measuring a properly rescaled normalized mutual information NMI
\begin{equation}
{\cal S}_{\text{NMI}}({\cal I},{\cal I}^{\text{NSR}})=\frac{ {\text{NMI}}({\cal I},{\cal I}^{\text{NSR}}) - {\text{NMI}}_\infty}{ 1-{\text{NMI}}_\infty},
\end{equation}
Results suggest that information extracted in IVG/IHVG is more robust to noise contamination than the raw information present in the image. This result is on agreement with previous works that showcase equivalent noise robustness results in VG/HVGs \cite{PRE, motifs}.

%Image visibility graphs can be used to assess the similarity between two images or the quality of a given image after a certain noise or filter is applied. Let's consider two Image Visibility Graphs of $N^2$ nodes extracted from two different images, ${\cal I_1}$ and ${\cal I_2}$, of the same size. Since the mapping of the image pixels into the corresponding IVG is unique we will have a one-to-one correspondence between the nodes of the two IVGs. Thus any structural measure able to assess the similarity of two graphs with the same set of nodes can be in principle used to measure how similar the two images are.\\
%The simplest measure one can think of is the node overlap(CITE), that is the percentage of links the two graphs have in common and is given by: 
%We used this index $Q$ to measure the similarity between the standard grayscale Lena image (see Figure\ref{fig:filters}) and the same image where we set to zero the value of a certain number of random pixels. In Figure\ref{fig:similarity} we show the values of similarity given by $Q$ in the case of IVG and IHVG, and we compare them with the respective values of the Normalized Mutual Information (CITE)(NMI) between the images, in function of the fraction $r$ of random pixels arbitrarily set to zero. 

\subsection{Visibility Filters}
\vspace{-2mm}

The results shown in Fig.\ref{fig:similarity} demonstrate that IVG/IHVGs encapsulate information which is robust against noise contamination. This virtue opens the possibility of using IVG/IHVGs to handle real images, which by construction are always polluted with noise. In this subsection we consider employing the IVG/IHVGs as filters.\\
%are robust encoders of structural information contained in images. This is a special property of the IHVG algorithm, which is based on the rank statistics of the pixels data of the image, and it makes IHVGs suitable as general image filters for several purposes such as image compression and image pre-processing.\\ 

\noindent {\bf Definition (Visibility filter). } {\it Let ${\cal I}$ be a $N\times N$ matrix where ${\cal I}_{ij}\in \mathbb{R}$. We define the visibility filtered image, or simply {\it visibility filter} ${\cal F}_{\text{VG}}({\cal I})$ as a $N\times N$ matrix whose $ij$ element corresponds to the value of a specific node property of the node $ij$ of the associated IVG, such as its degree or its clustering coefficient. If the property is a real scalar quantity, then this quantity is {opportunely rescaled to grayscale integer values}. The {\it horizontal visibility filter} ${\cal F}_{\text{HVG}}({\cal I})$ follows equivalently.}\\

The above definition of a visibility filter reduces to a particular topological plot with a trivial rescaling, which allows to compare across different node properties in comparable ranges (grayscale intensity range in $\{0,255\}$).
\begin{figure}
\includegraphics[width=1.05\columnwidth]{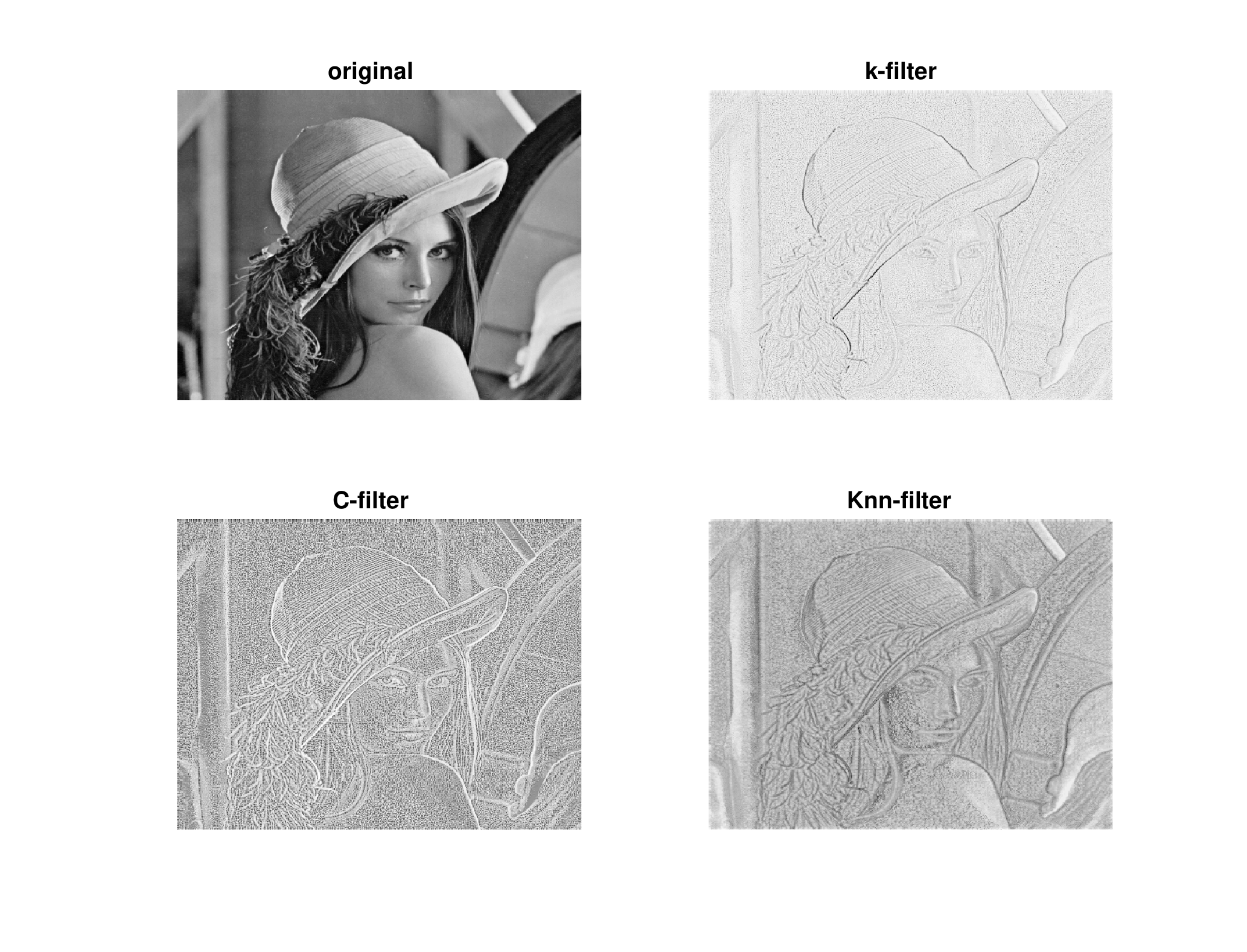}
\caption{Several Visibility Filters ${\cal F}_{\text{HVG}}$ applied on Lena.}
\label{fig:filters}
\end{figure}
A priori one can use many different node properties to define such filters: here we focus on three of the simplest properties, namely degree, clustering and degree-degree \cite{vito} to define three filters ${\cal F}_{\text{HVG}}({\cal I})$:\\
\begin{itemize}
\item {$k$-filter:} when we assign to $ij$ the degree of node $ij$. No rescaling is necessary here.
\item {$C$-filter:} when we assign to $ij$ the local clustering coefficient (rescaled in $\{0,255\}$) of node $ij$.
\item {$Knn$-filter:} when we assign to $ij$ the closest integer value from the average nearest-neighbors degree of node $ij$.
\end{itemize}

%\textcolor{red}{In principle in an IVG extracted from a gayscale image of M pixels k can take values in the interval [3, M-1], but the probability of having a node of degree k decreases exponentially fast with k (see Figure\ref{fig:compression} b)), thus no rescaling is necessary to construct a k-filter}.
In Fig. \ref{fig:filters} we plot the original grayscale Lena image ($512\times512$) together with the three filters in a reverse grayscale. By observing these filtered images it's easy to notice that many important details (for example details in the Lena face and in the hat) are still recognizable, but the information from the original image has been certainly compressed.
%Note that, visually, these filters share some similarities with edge detection filters (e.g. Sobol, Canny, entropy filters).
%While important aspects of the image seem to be encapsulated in the filtered images, a large amount of information seems to have been swept away, 
%suggesting good compression properties of these filters. 
Accordingly, we briefly explore compression properties of these filters in the next subsection.  

\subsection{Compression}
\vspace{-2mm}
In grayscale, a generic squared image ${\cal I}$ has $N^2$ pixels and each pixel can take $2^8$ values, so the image size is $|{\cal I}|=N^2$ bytes. In the IHVG $k$-filter, each pixel corresponds now to the degree $k$ of the associated node. We know that by construction in HVGs, neglecting boundary effects, we have $8\leq k\leq N^2-1$. Now, there is a theorem that guarantees that for white noise images, the degree distribution of ${\cal F}_{\text{HVG}}({\cal I})$ decays exponentially \cite{scalar}, so the probability of finding a degree larger than $K$ is
$$P(k>K)=\int_{K}^\infty \bigg(\frac{1}{9}\bigg)\bigg(\frac{8}{9}\bigg)^{k-8}dk=9^{7-K}\cdot 8^{K-8}$$
To give an idea of the order of magnitude, for $K\approx 24$, this probability is lower than 0.015, i.e. less than $1.5\%$ of the nodes have a degree larger than 24. In other words, if we decided to cut-off information of the degree and assign $k=K$ for all nodes whose degree is larger or equal to $K$, we would only cut-off less than  $1.5\%$ of the nodes in the $k$-filter of a white noise image. Now, at least in the 1D (time series) version it is known that the degree distribution of HVGs also decays exponentially in the typical case (not just for white noise), so we expect that a similar behaviour takes place in the image setting, and thus the probability of finding large degrees is, generically, exponentially small. This actually explains why no rescaling in practice is necessary to construct a $k$-filter.\\
If no cut-off is set on the degree, there is an exponentially vanishing (although not zero) probability of finding that a node gets degree $k$, for any $k<N^2-1$. In practice, by setting a cut-off value (for instance set $k_{\text{cut-off}}=24$) such that we assign the value $k_{\text{cut-off}}$ to all the pixels whose actual degree is either $k_{\text{cut-off}}$ or larger,  by construction,
$$|{\cal F}_{\text{HVG}}({\cal I})|=\frac{\log_2(k_{\text{cut-off}} - 8)}{8}N^2 \ {\text{bytes}}.$$
Note that for $k_{\text{cut-off}}=24$ we only need $N^2/2$ bytes, so we have a $50\%$ size reduction with minimal degradation.\\ 

\begin{figure}
\includegraphics[width=1\columnwidth]{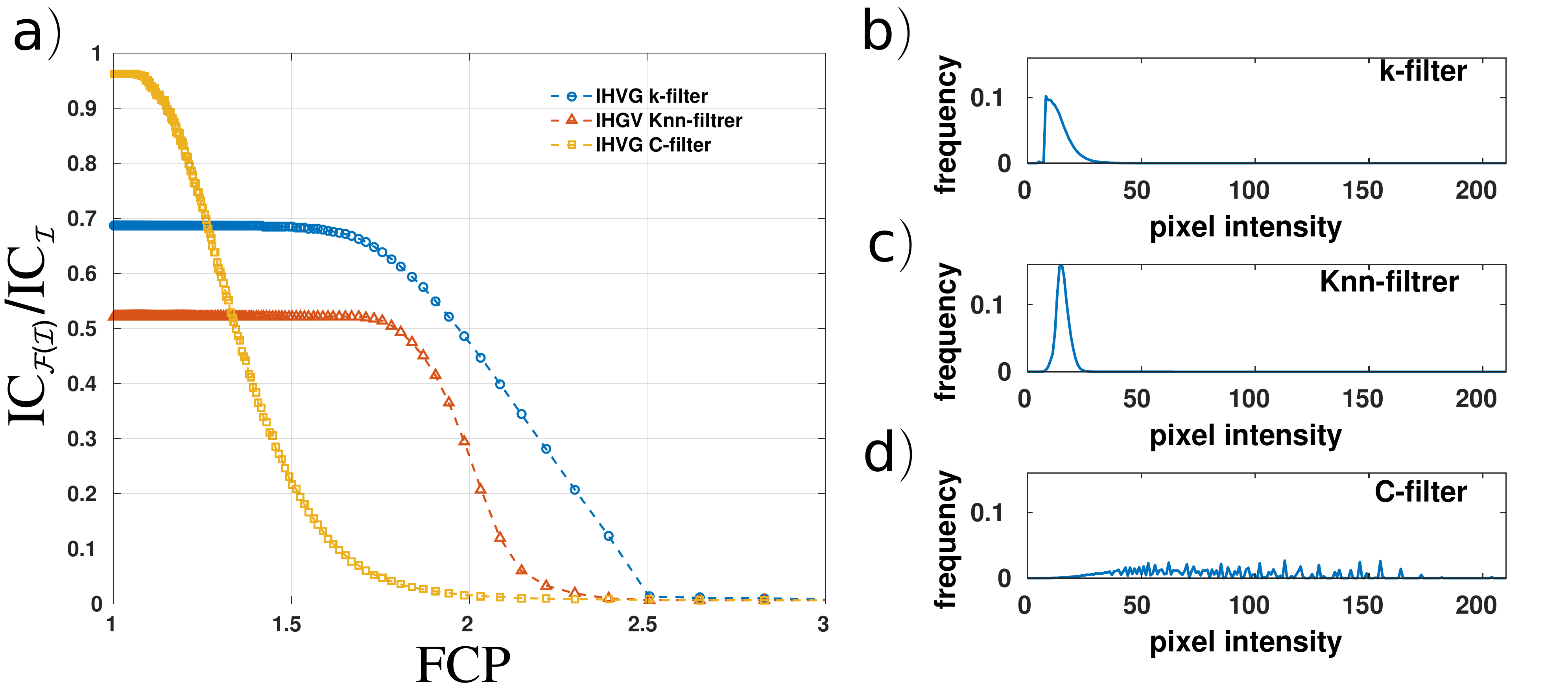}
\caption{{IHVG filters applied to the Lena image for testing the compression performance. Panel a) The percentage of image complexity $\text{IC}_{\cal F(I)}/\text{IC}_{\cal I}$ captured by the filtered image is plotted as a function of the (theoretical) compression level FCA=$\log_2(\max\{{\cal I}_{ij}\}+1)/\log_2(x_{\text{cut-off}}+1)$, which can be tuned by setting a cut-off value $x_{\text{cut-off}}$ on the pixel-intensity of the filtered images. Panel b) The pixel-intensity distribution for the IHVG $k$-filter is peaked and decays exponentially, enabling to reach high level of compression while still encoding consistent amount of complexity. Panel c) The IHVG $Knn$-filter pixel distribution also show an exponential decay in the pixel-intensity distribution but its performance is worst than the one of the $k$-filter. Panel d) In turn, the IHVG $C$-filter is characterized by a broad distribution of pixel intensities which makes this filter the best encoder of information only when little or no compression is required.} } 
\label{fig:compression}
\end{figure}

\begin{table*}[]
\centering
\begin{tabular}{|l|l|l|l|l|l|l|l|l|}
\hline
                    & FCP  & FCP$^{\alpha=.05}$   & s(${\cal I}$), s(${\cal F(I)}$)    & $s({\cal C}({\cal I}))$    & $s({\cal C}({\cal F(I)}))$    & $\text{IC}_{\cal I}$ & $\text{IC}_{\cal FI}$ & $\text{IC}_{\cal F(I)}/\text{IC} _{\cal I}$ \\ \hline
Lena            & 1.24 & 1.73 & 263.2 kB & 223.9 kB & 153.8 kB & 0.85  & 0.58   & 0.69          \\ \hline
cameraman       & 1.41 & 1.82 & 263.2 kB & 161.8 kB & 136.3 kB & 0.62  & 0.52   & 0.84          \\ \hline
pirate          & 1.3 & 1.73 &263.2 kB & 221.6 kB & 154.8 kB & 0.84  & 0.59   & 0.7           \\ \hline
peppers         & 1.33 & 1.8 & 263.2 kB & 143.8 kB & 137 kB   & 0.55  & 0.52   & 0.95          \\ \hline
house           & 1.45 & 1.82 & 263.2 kB & 129.9 kB & 103 kB   & 0.49  & 0.39   & 0.79          \\ \hline
mandryl         & 1.33 & 1.71 & 263.2 kB & 231.9 kB & 147.8 kB & 0.88  & 0.56   & 0.63          \\ \hline
livingroom      & 1.28 & 1.74 & 263.2 kB & 221 kB   & 155.1 kB & 0.84  & 0.59   & 0.7           \\ \hline
\end{tabular}
\caption{Measures to evaluate IHVG $k$-filter compressor performance on different standard grayscale test images reported in Figure \ref{fig:testimages}. FCP, filter compressing power; FCP$^{\alpha=.05}$, optimal filter compressing power; s(${\cal I}$)(s(${\cal F(I)}$)) of the original(filtered) image in bitmap format; $s({\cal C}({\cal I}))$ and $s({\cal C}({\cal F(I)}))$ are the sizes of the zipped original image and of the zipped filtered image respectively; $\text{IC}_{\cal I}$($\text{IC}_{\cal FI}$) image complexity of the original(filtered) image.}
\label{tab:compression}
\end{table*}

{In order to make a systematic analysis of the different filters, we define the size of the original Lena image as 
$$|{\cal I}|=\frac{\log_2(\max\{{\cal I}_{ij}\}+1)}{8}N^2$$
where in our case $\max\{{\cal I}_{ij}\}=245<2^8$,
and the size of the filtered Lena images ${\cal F}_{\text{HVG}}(I)$, or simply ${\cal F}({\cal I})$ as
$$|{\cal F}({\cal I})|=\frac{\log_2(x_{\text{cut-off}}+1)}{8}N^2,$$
where $x_{\text{cut-off}}$ is a node-value cut-off in the filtered image. Incidentally, note that in IHVGs the minimum degree is $k=8$, so the numerator in the preceding equation should read $\log_2(k_{\text{cut-off}} - 8)$ instead of $\log_2(k_{\text{cut-off}} +1)$, however for generality we stick with the definition above (not all filters have lower bound values) and only note that, in what follows, results for the case of the degree filter are being rather conservative.\\
%\textcolor{blue}{We are being conservative! Note that for the degree, we don't need to allocate memory for the numbers 1 to 8 as we know that the degree is always larger or equal to 8, but this is not subtracted in the formula. That is not clear anymore in the C-filter of Knn-filter, but we need to think a way of expressing this for the degree. Something like a $x_{min}$.} \textcolor{red}{Ok but this argument is not general for all filters. At the end of the section we can say something like 'note that in the case of the degree filter we have been conservative because by applying a pad to the images which set the value $x_{min}=8$ (see equation at beginning of section) for the pixels at the borders we can further reduce the size of the filter without losing any information.}
This in turn enables us to define a general {\it{filter compression power}} (FCP) as the ratio  
$$\text{FCP}=|{\cal I}|/|{\cal F}_{\text{HVG}}({\cal I})|.$$ 
%The filter compression ratio so defined is a normalized quantity that expresses the fraction of information bytes in the original image that are unnecessary to store the filtered image.
When no cut-off is enabled on to the filter, then $x_{\text{cut-off}}=x_{\text{max}}$, and the corresponding value of FCP reduces to the {\it{natural}} compressing power of the filter.}
\begin{figure}
\includegraphics[width=0.8\columnwidth]{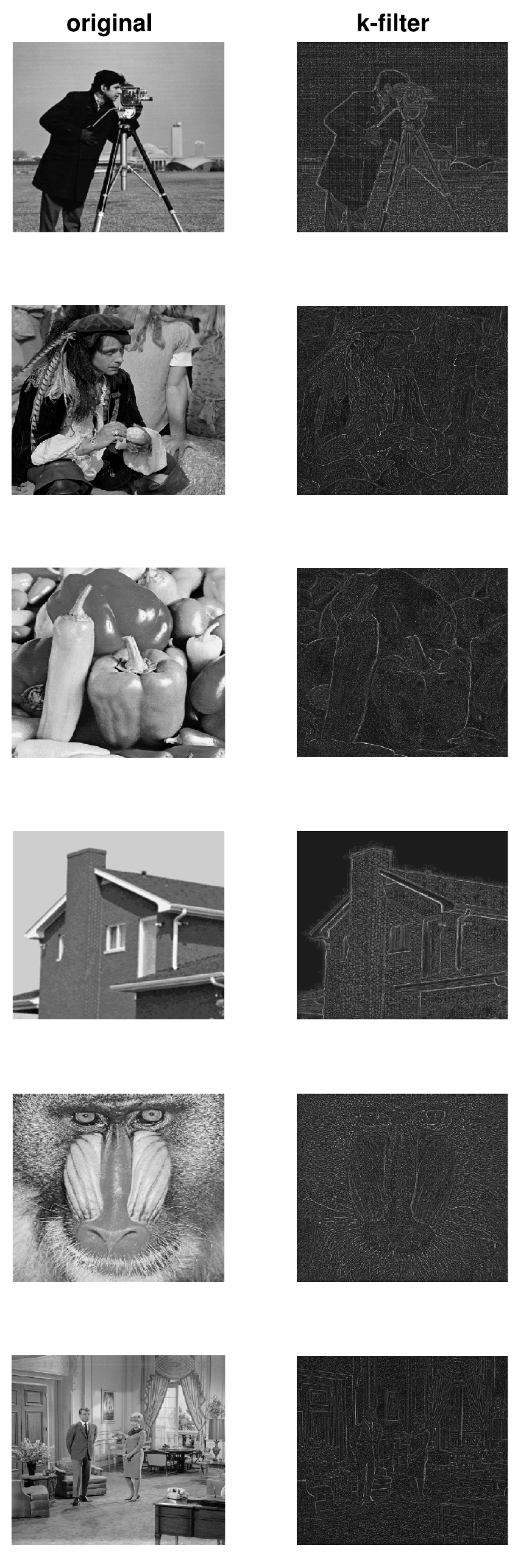}
\caption{{The set of 512x512 standard grayscale images often found in literature that has been used to test the compressor performance of the IHVG $k$-filter.}}
\label{fig:testimages}
\end{figure}
By lowering the value of $x_{\text{cut-off}}$ below $x_{\text{max}}$ we can further increase the effective compression power at the expenses of deliberately degrading the filtered image. To study how much information from the original image the IVG/IHVG filters are still able to encode given any arbitrary level of compression ratio, we now need to use a measure of spatial information.  
To this aim we shall compute an {\it image complexity} (IC) which can be estimated using standard compressor (zip) algorithms, such that
\begin{equation}
\text{IC}_{\cal I}=\frac{s({\cal C(I)})}{s({\cal I})} 
\end{equation}
where $s({\cal I})$ is the measured size in kB of the image $\cal I$ given by the computer machine  after saving the images in bitmap format, and $s({\cal C}({\cal I}))$ is the measured size in kB of the zipped file of the image compressed using a Lempel-Ziv-type compressing algorithm. All over this work we use \textsc{Matlab}'s zip routine.\\

\noindent Accordingly, we can then quantify how much image complexity of the original image is encapsulated by different filters at any desired level of compression power. In Figure \ref{fig:compression} (panel a) we plot the ratio $\text{IC}_{\cal F(I)}/\text{IC}_{\cal I}$ as a function of FCP, for the three IHVG filters under study for a range of cut-off values %$x_{\text{cut-off}} \in \{255,254,...,0\}$ 
(a similar analysis can be performed on IVG, data not shown). Increasing values of FCP corresponds to lowering the value of the cut-off. In Figure \ref{fig:compression} b),c) and d) the distribution of the node-values for the different filters (degree, clustering and degree-degree) are reported to better understand the curves. 
If no or very little compression is necessary, we find that the $C$-filter (for which the clustering distribution is broad) almost encapsulates all the original image complexity ($\text{IC}_{\cal F(I)}/\text{IC}_{\cal I} \simeq$ 0.97 for FCP$\simeq$1) and performs better than the $k$-filter and of the $Knn$-filter. The drawback of this filter is that the information it encapsulates is rapidly degraded when one wishes to achieve even a modest level of compression ($\text{IC}_{\cal F(I)}/\text{IC}_{\cal I} \sim$ 0.2 for FCP=1.5). On the other hand,  the $k$-filter (whose degree distribution is peaked and exponentially decaying) is able to still encapsulate high content of information even when a high level of compression is set ($\text{IC}_{\cal F(I)}/\text{IC}_{\cal I} \simeq$ 0.5 at FCP=2).
The curves also suggest that it is possible to improve consistently the $k$-filter natural compressing power by getting rid of a very little fraction of the spatial information encoded by the filter.\\

%\noindent \textcolor{blue}{I dont get this totally}.\textcolor{red}{ Look at TableII: when you filter Lena without any x-cutoff you have FCP=1.24 and a certain amount of complexity captured. Now look at figure 7. You have a nice plateau. So what I'm saying is that it's convenient to give up another (small amount) 5 percent of the complexity the filter is capturing by applying an x-cutoff to the filter. In this way you gain a lot in FCP (now FCP$^{.05}$=1.75) and you lose a negligible amount of complexity}. 
\noindent According to these results, we can now define a {\it{filter optimal compressing power}} (FCP$^{\alpha}$) for the value of the cut-off $x_{\text{cut-off}}^{\alpha}$ for which 
$$x_{\text{cut-off}}^{\alpha} : \frac{\text{IC}_{{\cal(F(I))}_{x_{\text{max}}}}-\text{IC}_{{\cal(F(I))}_{x_{\text{cut-off}}^{\alpha}}}}{\text{IC}_{\cal I}}=\alpha$$
where $\alpha$ is a parameter which measure the maximum tolerance on the amount of information degraded by the filter and has to be small.\\ 
Here we set $\alpha$=0.05 and we test the performance of the $k$-filter as a compressor on several $512\times512$ standard grayscale images (the images can be found in the Image Database website\footnote{\texttt{http://imageprocessingplace.com}} and are plotted together with their corresponding $k$-filtered images in Fig. \ref{fig:testimages}). Results are summarised in Table \ref{tab:compression}.

\subsection{IHVG filters for image pre-processing}

Consider a standard problem in image processing: visual face recognition. The canonical approach consists in extracting from each image a feature vector and subsequently feed classifiers with these vectors, for supervised learning.
In some cases the performance of the classifier can be improved if images are pre-processed (i.e. before feature extraction) \cite{nanni}. For example local binary patterns (LPB) \cite{lbp,lbp2,lbp3} that are robust to monotonic grayscale variations are widely used in face recognition problems to overcome the fact that often images are taken in different illumination conditions or with different resolutions.\\
 
\begin{figure}[h!]
\includegraphics[width=1.0\columnwidth]{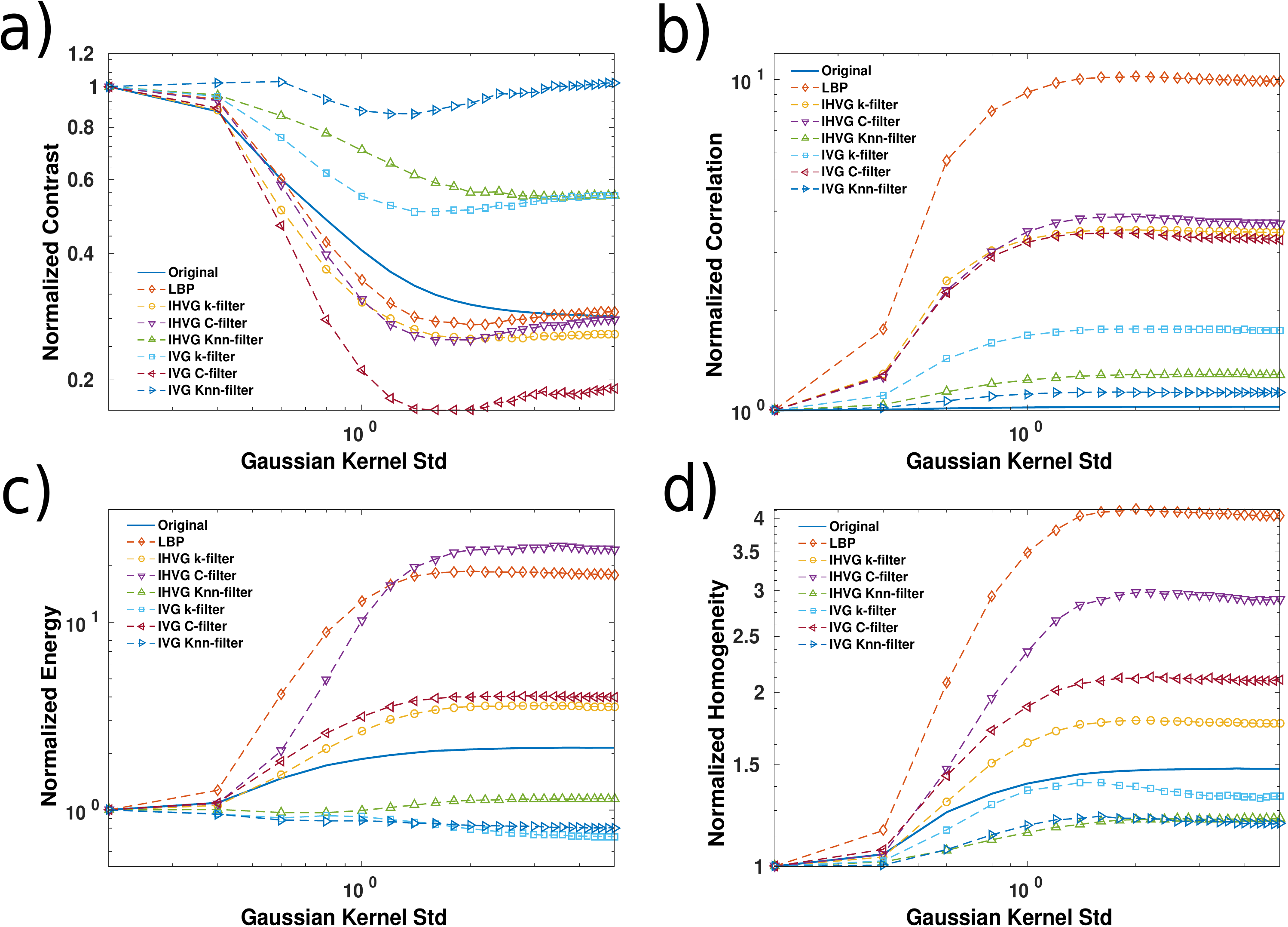}
\caption{{Haralick features Contrast, Correlation, Energy and Homogeneity extracted from the lena image when Gaussian kernels of different variance are applied to it and different IHVG/IVH filters are used to pre-process the image before the extraction. The features are all normalized (rescaled) with respect to the value of the original image (Gaussian kernel std=0).}}
\label{fig:haralick}
\end{figure}

\noindent Here we test the potential of the different IVG/IHVG filters defined above as pre-processing filters for the improvement of image texture analysis \cite{nanni}. To simulate real-world acquisition variability we transform an image of Lena to different resolution scales by applying to the original image  a $5\times5$ Gaussian kernel with increasing values of standard deviation. As a standard feature extraction method we then make use of the gray level co-occurrence matrix (GLCM), \cite{haralick,nanni}  a well known texture analysis technique. 
%to extract descriptors from the image of Lena's face at different resolution scales, to simulate real-world acquisition variability. To practically simulate different resolution scales of the image we apply a $5\times5$ Gaussian kernel with increasing values of standard deviation.\\
For a given resolution (kernel std value) the GLCM is computed from the image and four so-called Haralick features are extracted: Contrast (measuring local variations), Correlation (measuring joint probability occurrence of pairs), Energy (sum of squared elements of the matrix) and Homogeneity (measuring the closeness of the distribution of elements in the matrix to its diagonal), see \cite{haralick} for details. 
%Subsequently,  the $L_1$ distance between each of these four descriptors and the corresponding descriptors from the original image at the best resolution are measured, and an overall distance $\cal D$ is obtained by summing up all the individual distances.\textcolor{blue}{You need to give more details here: why $\cal D$ is a distance? In which way $\cal D$ is related to classification accuracy? CITE} The function $\cal D$ of the resolution is indeed a general classification performance indicator for any classifiers trained on the four selected features. 
All images are then mapped into points in the 4-dimensional space spanned by the appropriately normalized Haralick features {(see Fig. \ref{fig:haralick})}. The distance $\cal D$ (using city-block ($L_1$) norm, results hold similar for an Euclidean metric) between low resolution images and the original one is subsequently computed. Intuitively, the lower $\cal D$, the more similar a low-resolution image is from its high-resolution version in feature space, and accordingly a classifier trained with the high-resolution version would a priori find less difficulties to correctly classify low-resolution images for the filter with overall lower values of $\cal D$.\\
%It is reasonable to assume that a simple image classifier taking in input Haralick features will produce a similarity value between any two images based on a metric-dependent distance. 
%In other words, $\cal D$ can be considered as an indicator of the classification performance of a simple classifier trained on the four Haralick selected features (the lower $\cal D$, the higher the accuracy).\\

\begin{figure}[h!]
\includegraphics[width=1.0\columnwidth]{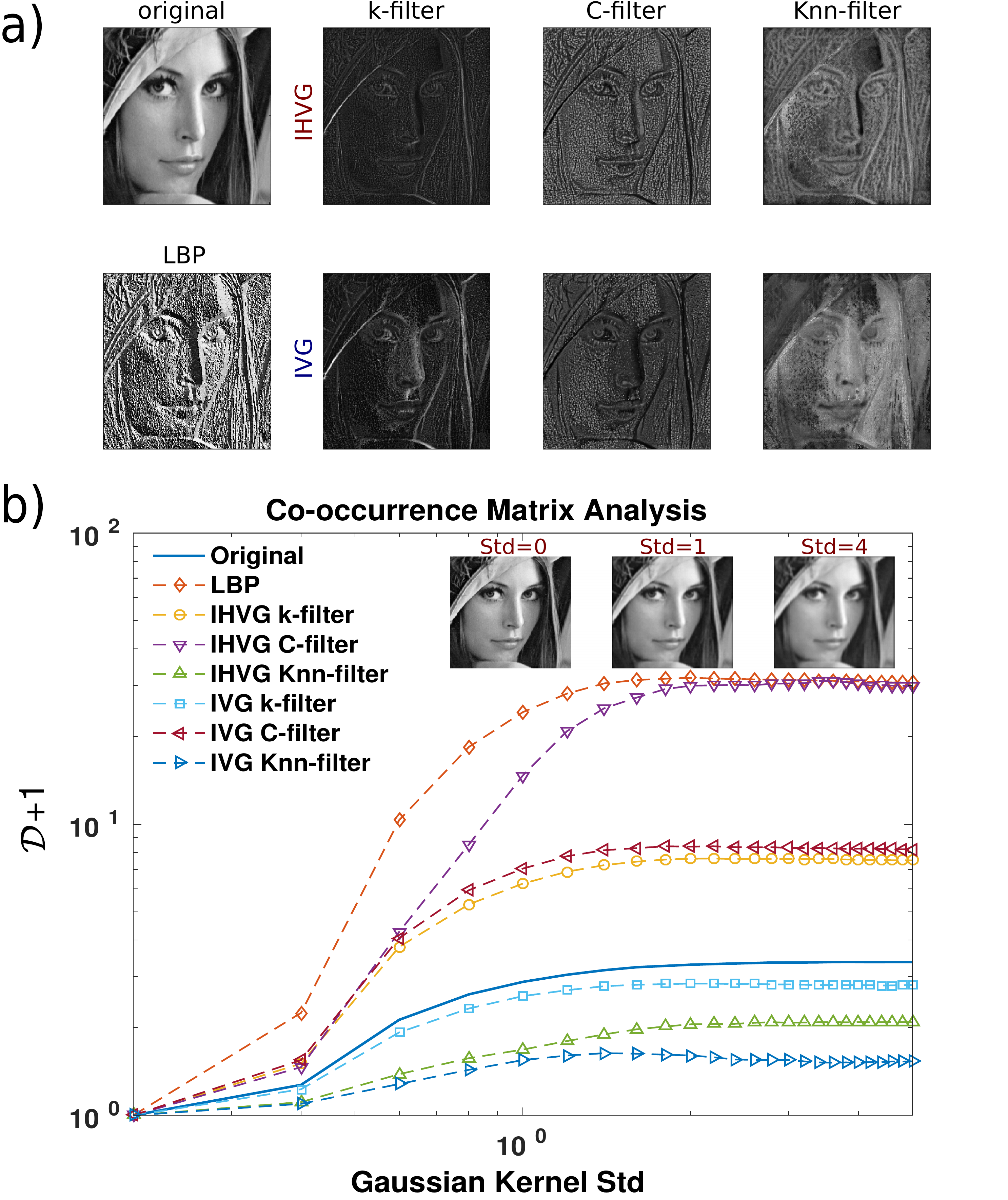}
\caption{{Performance of the IVG/IHVG filters in image pre-processing. a) The IVG and IHVG filters are shown together with the LBP (local binary patterns) filter as applied to the Lena face image. b) The distance 
$\cal D$ between the Lena face image and the same image at different resolution scales is measured in the 4-dimensional space of the {(normalized)} Haralick features Contrast, Correlation, Energy and Homogeneity and plotted in function of the value of standard deviation of the Gaussian 5x5 kernel applied to obtain low-resolution images. The distance curve is computed for the original unfiltered image as well as for the same image after being pre-processed using the filters shown in a). $\cal D$ is an indicator the performance of the filters in pre-processing images to improve feature extraction.}}
\label{fig:preprocessing}
\end{figure}

\noindent The procedure is repeated several times for (i) the non-filtered image, and for (ii) filtered images using all six visibility filters ($k$-filter, $Knn-$filter, and $C-$filter for IVG and IHVG); and comparison with the LBP filter is reported as well for completeness. Results are shown in Fig. \ref{fig:preprocessing} b), where we plot in log-log ${\cal D}+1$ as a function of the kernel standard deviation (i.e. the larger this parameter, the lower the image's resolution). In panel a) of the same Figure, the filter effect for the different filters used in the analysis on the Lena face is shown for illustration. We find that in the case of both IVG and IHVG $Knn$-filters and the IVG $k$-filter, $\cal D$ appears systematically below the one for the non-filtered image, suggesting that all low-resolution images will be systematically closer (in feature space) to the high-resolution version after these specific filtering pre-processing. 

%\textcolor{blue}{A referee will probably tell us: why dont you do the classification and check whether that's true? i.e. take say 5 different figures, each of them with a lot of samples with different resolutions, and classify them, and then compare the results with the same procedure applied on the filtered images. Not saying we need to do more work, just saying the referees will ask that with high probability (I would).} \textcolor{red}{I agree but I also like the idea of showing the test from a more theoretical perspective by using D.}

%\textcolor{red}{}

\section{Pattern Recognition and Texture Classification}
%\textcolor{red}{!!!THIS SECTION IS WORK-IN-PROGRESS!!!}
Once we have discussed the usefulness of image visibility graphs for filtering and compression purposes, we now turn and explore their applicability as efficient and universal feature-extraction methods for the task of texture detection and classification, a classical problem in computer vision \cite{advancetexture,computervision}.\\ 
Despite no rigorous mathematical definition of texture is widely accepted, textures can be intuitively defined as characteristic visual patterns arising from the spatial distribution of the pixel intensity values in images \cite{multilayerCN}, and therefore they are of fundamental importance in many machine vision applications such as object tracking, face recognition and automated medical diagnosis \cite{reviewtexture2}.
Following recent efforts in the direction of cataloguing textures and texture-related datasets \cite{reviewtexture,reviewtexture2}, we will test the performance of image visibility related descriptors on different types of textures, namely textures of materials, bio-medical textures and natural textures.

\begin{figure}[h!]
\includegraphics[width=1.0\columnwidth]{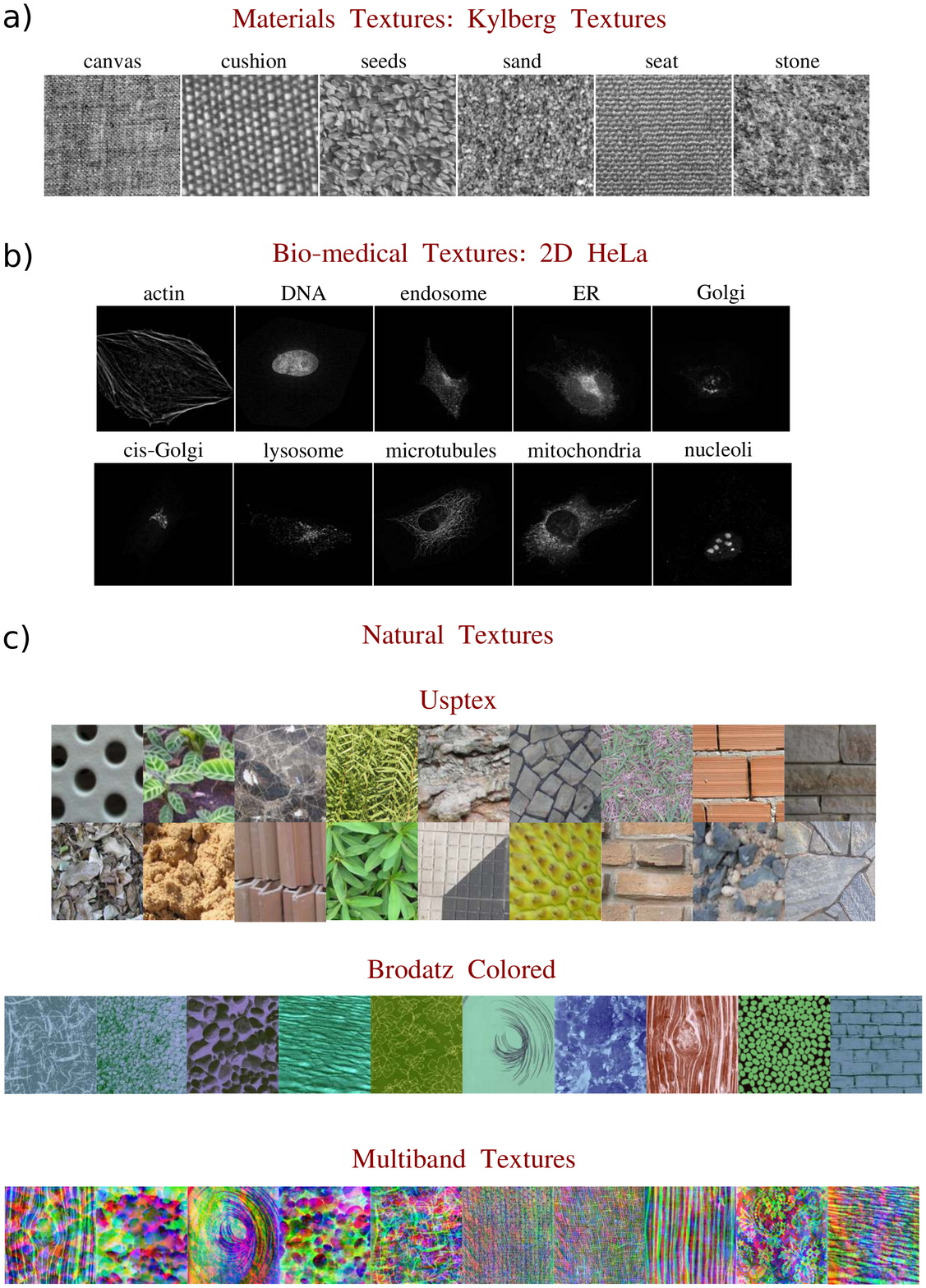}
\caption{(Color online). Datasets used to validate the usefulness of image visibility features in texture recognition: a) the Kylbert reduced dataset includes grayscale images of texture from six different materials. b) 2D HeLa contains grayscale grayscale fluorescence microscopy images of ten organelle-specific proteins. c) Three dataset are used for the task of natural texture classification: the Upstex dataset, the Brodatz Colored dataset, and the Multiband dataset (MTB). In the Upstex dataset each image is a sample from of a class (191 classes in total, 12 images per class), while for Brodatz and MTB each image shown include all the images of a class and its 16 non-overlapping sub-images represent the samples (151 classes in Brodatz, 154 classes in MTB).}
\label{fig:datasets}
\end{figure}

\begin{table*}[tb]
\centering
\begin{tabular}{|c|c|c|c|c|c|c|}
\hline
& Complex Tree & Linear Discriminant & linear SVM & quadratic SVM & weighted KNN & Ensemble Bagged Tree \\ \hline
IHVG P(k)  & {\bf100}\% & 96.7\%  & {\bf 100}\% & {\bf 100}\%  & 98.8\%    &{ \bf 100}\%   \\ \hline
IHVG P(k)+PCA (5 PCs) & {\bf100}\% & 96.7\%  & {\bf 100}\% & {\bf 100}\% & {\bf 100}\%  & 99.2\% \\ \hline
IVG P(k)  & 95.3\%   & 95.8\% & 99.2\%   & {\bf 100}\%   & 94.2\%   & 98.8\%  \\ \hline
IVG P(k)+PCA (5 PCs) & 96.7\% & 97\% & 97.9\% & {\bf 100}\%  & {\bf 100}\%  & 98.8\% \\

\hline   
\end{tabular}
\caption{Kylberg Texture Dataset: performance in classification accuracy obtained with different algorithms. The classifiers used are standard classifiers implemented in the \textsc{Matlab} Classification Learner App. }
\label{tab:1}
\end{table*}
\subsection{Global features for Materials Texture Recognition}

\begin{figure}[h!]
\includegraphics[width=1.\columnwidth]{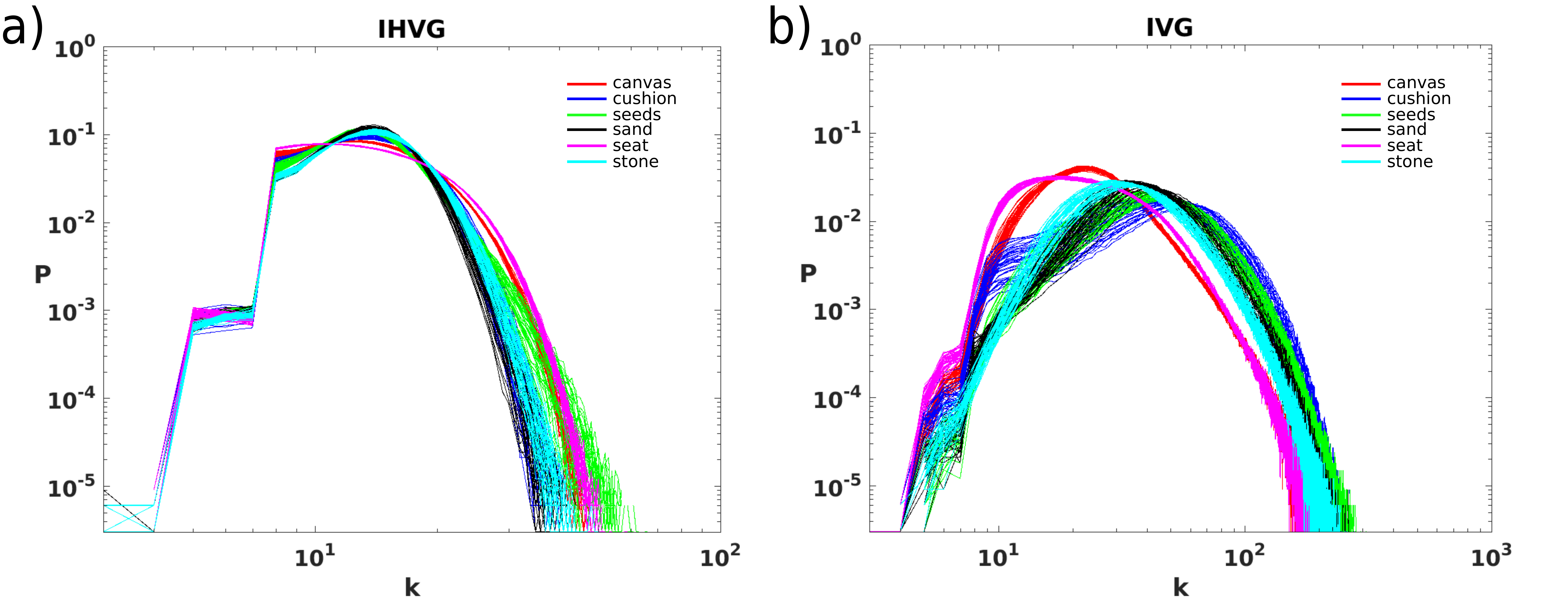}
\caption{(color online) Log-log plot of the degree distributions $P(k)$ for a) all IHVGs and for b) all IVGs extracted from the 6 classes of the Kylberg Texture Dataset. This global features appears to be capturing a great deal of the differences between textures, reaching a $100\%$ classification accuracy using a linear Support Vector Machine (see table \ref{tab:1}). %\textcolor{red}{more comments: 1)100\% for several classifiers PCA 2) PCA still capture big deal and helps KNN c) IHVG global seems to perform better than INVG}
}
\label{fig:ihvgPk}
\end{figure}

We start by showing that the degree distribution $P(k)$ extracted from IHVGs is an ideal global feature vector for the automatic recognition of {\it material textures}.
These are a specific class of textures derived by imaging the surface of various types of materials for the purpose of classifying them according to the characteristic structural patterns captured in this way. The dataset we used is a reduced version of the Kylberg Texture Dataset \cite{kylberg}, formed by grayscale texture images from 6 different 
materials (canvas surface, cushion surface, line seeds, sand, seat surface and stone surface) with 40 samples per class. 
This dataset is often used as a benchmark for testing material texture descriptors because of the image homogeneity in terms of perspective, scale, and illumination across the different classes \cite{kylberg2,andrearczyk}. Because of the grayscale nature of the images and the uniform texture of the materials, this is ideal to test the capability of IVG/IHVGs global features to capture the surface structural correlations and patterns. 
In Figure \ref{fig:datasets} a) we provide an illustration of the texture classes.
Each image in the dataset is transformed to the corresponding IHVG and its degree distribution $P(k)$ is extracted and then given in input (as a vector of features) to different classifiers. Principal component analysis (PCA) is applied to the input vectors to reduce the number of features and to avoid eventual overfitting, and a $5-$fold cross validation procedure is adopted.  A $100\%$ (average by class) classification accuracy --defined as the percentage of samples correctly assigned to their true class-- can be obtained by mean of a linear/quadratic-kernel Multiclass one-vs-one Support Vector Machine (ovo-lSVM/qSVM), for IHVG and with a quadratic-kernel SVM for IVG. Interestingly, this level of accuracy is already obtained only using the first 5 principal components of the PCA projection. To further understand why the degree distribution is such an informative feature, in Fig. \ref{fig:ihvgPk} we plot the degree distribution for all the samples (colored by class as a guide for the eye) for the IHVGs (a) and for the IVGs (b). Distributions cluster by class and already by visual inspection we can determine the clusters. For completeness, in Table \ref{tab:1} we report the classification accuracies reached by different standard classifiers implemented in the \textsc{Matlab} Classification Learner App. For IHVGs it is possible to obtain an accuracy of $100\%$ by using a Complex Tree classifier or a Bagged Tree and with a weighted KNN algorithm as well after applying PCA. 
%We conclude that, in the context of texture recognition, global features from IHVGs are highly informative and seem to be capturing more information than those extracted from IVGs.

\subsection{Combining Global and Local Features for Bio-medical Texture Classification}

In a second step, we now focus on a more complex image dataset called 2D HeLa. This is a bio-medical dataset of grayscale fluorescence microscopy images of HeLa cells stained 
with various organelle-specific fluorescent dyes able to reveal characteristic structural patterns of proteins so that 10 classes can be identified: Actin, DNA, Endosome, ER, cis-Golgi, Golgi, Lysosome, Microtubules, Mitochondria and Nucleoli. For each of these classes there is a different number of sample images (between 73 and 98), for a total of 862 images. In Fig. \ref{fig:datasets} (b) we illustrate one representative sample for each class.This dataset is exemplar with respect to the bio-medical challenge of the automated interpretation of sub-cellular protein patterns in fluorescence microscope images, which can help characterize many genes whose function is still unknown.\\ 
Unlike material textures, these images show several complex biological structures that occupy only a certain subregion of the image. This fact precludes using only global features, justifying the necessity to complement these with local features to attain high classification accuracies. Accordingly, we test the performance of the Visibility Patches. \\ 

\begin{figure}[h!]
\includegraphics[width=1.0\columnwidth]{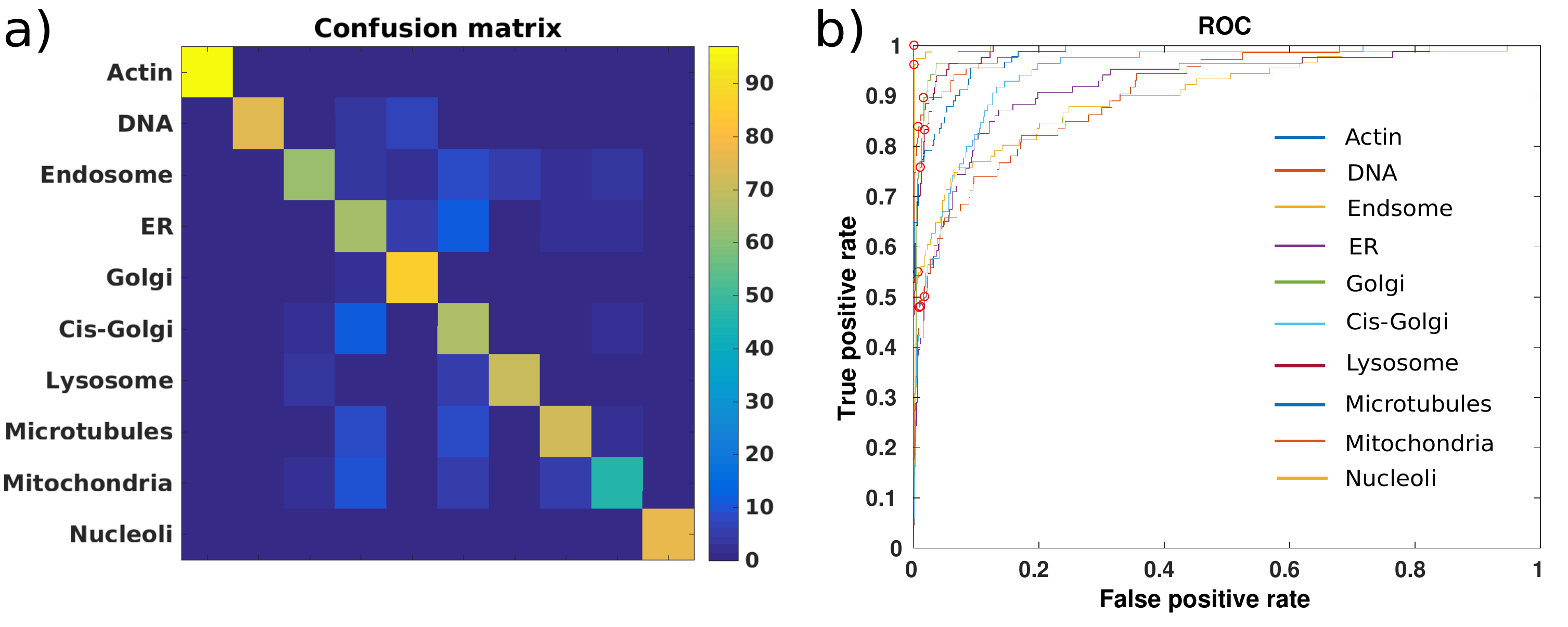}
\caption{Confusion matrix (panel a) and AUCs (panel b) for the classification of the 2D HeLa Dataset, using a quadratic kernel Support Vector Machine on a blend of Global (degree distribution) and Local (Visibility Patches) from IHVG and IVG.}
\label{fig:2DHeLa}
\end{figure}

\begin{figure}[h!]
\includegraphics[width=1.0\columnwidth]{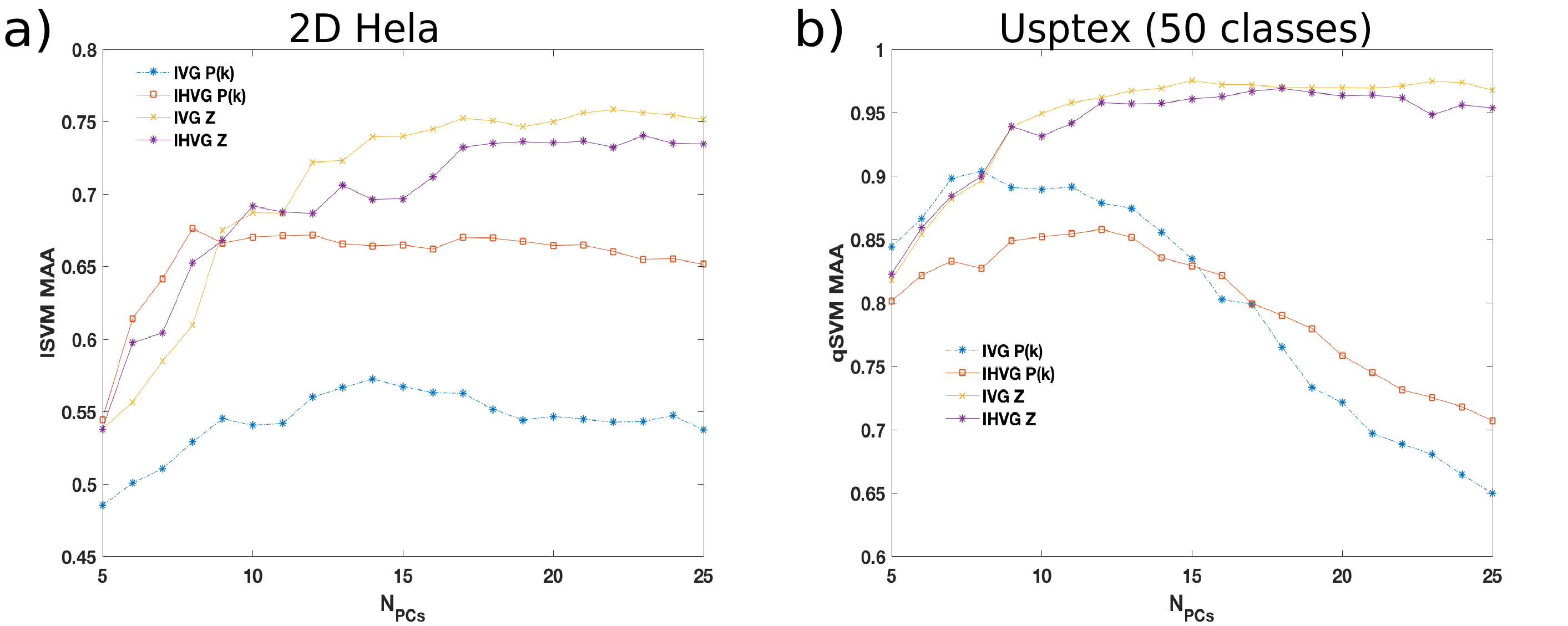}
\caption{Model average classification accuracy (30 model realizations) in function of the number of principal components ${\bf N_{PCs}}$ obtained with basic global/local descriptors using a) a linear-kernel Multiclass one-vs-one Support Vector Machine on the 2DHela dataset and b) a quadratic-kernel Multiclass one-vs-all Support Vector Machine on a subset of Usptext where only the first 50 classes are included. }
\label{fig:tuningPCs}
\end{figure}

\noindent Again, we feed the feature vector with global descriptors given by the first entries of the degree distribution $P(k)$ of IHVGs (up to $k=40$), the degree distribution of IVGs (up to $k=80$), and a concatenation of both. Additionally, we add local features given by the visibility patch profiles $\bf Z$ of HVP$_3$ (188 patches detected) and VP$_3$  (249 patches detected), and a concatenation of the two. To avoid overfitting, PCA is applied to reduce the number of features before classification. The PCA-transformed features are then given in input again to a linear-kernel Multiclass one-vs-one Support Vector Machine. 
%\textcolor{red}{WOULD BE GOOD WE REPORT THE FEW LINES OF MATLAB CODE?} 
The best classification accuracy is obtained by using a combination of global and local features from the blend of IHVG and IVG descriptors, reaching 83.5\%. 
In Table \ref{tab:2} we report, for each set of descriptors, the values of the best average classification accuracy and corresponding average (over the classes) area under the ROC curve (AUC)  obtained over 30 realizations of the classifier, together with the average classification accuracy over of the model realizations (Model Average Accuracy MAA). In Fig. \ref{fig:2DHeLa} we depict the details of the confusion matrix and of the ROC curves by class for the best model trained. The highest off-diagonal values in the confusion matrix correspond to pairs of classes that are expected to be hard to distinguish, for example ER and mitochondrial proteins and Golgi and cis-Golgi proteins.
The optimal number of PCs for each feature set (local/global, IVG/IHVG) is found with an exhaustive search in the space of the principal components number ${\bf N_{PCs}}\in[5,...,25]$ by measuring the average classification accuracy of the model (MAA) over 30 model realizations (see Figure \ref{fig:tuningPCs} (a) and Table \ref{tab:2}). The values of MAA for each descriptor tend to quickly saturate by increasing the number of components, meaning that the final model is robust with respect to the choice of the set of ${\bf N_{PCs}}$). 
Interestingly, we observe that while global features from the IHVG are more informative than for the IVG, the opposite is true for local features, as these seem to be more informative for IVG than for IHVG.  For comparison, in the original paper \cite{2dhela} of the 2D HeLa dataset the authors obtained an average class accuracy of 84\% by training a Back-Propagation Neural Network with specifically designed (ad-hoc) features. In other words, our parsimonious approach, using a set of non-specific graph-based features is competitive with respect to ad-hoc, context-dependent features. The highest published performance we are aware of is 95.3\% \cite{2dhela2}, obtained with a multi-resolution feature extraction approach combined with a neural network classifier, and it is an open problem whether one could reach competitive results by complementing the non-specific approach designed here with problem-specific strategies.\\

%\textcolor{blue}{While our results are arguably good, they are overcome by more sophisticated methodologies, something expected taking into account that the set of features employed here are non-specific.}\\
%\textcolor{red}{I won't say that. First our features being non-specific are more general, which is better. Second the revised results show that we are competitive with ad-hoc features. The problem is that or classification workflow is old-school: features-PCA-SVM...we are not 'deep' in a learning sense. I would like to give a feeling to the reader that this work is somehow focused on the feature definition and quality assessment but still there are many ways to be explored to implement the use of IVG features in classifiers with super-fancy architectures used today.}

\begin{table}[]
\centering
\begin{tabular}{|l|l|l|l|}
\hline
ovo-lSVM models                                                       & Accuracy & $\langle \text{AUC}\rangle$     & MAA   \\ \hline\begin{tabular}[c]{@{}l@{}}{\it Global features:} P(k)+PCA\\ \quad\quad\quad  IVG+IHVG\end{tabular} & 75.4\%    & 0.939   & 74.5\% \\
\begin{tabular}[c]{@{}l@{}}{\it Local features:} {\bf Z} + PCA\\ \quad\quad\quad   IVG+IHVG\end{tabular}  & 79.5\%    & 0.961  & 78.6\% \\
\begin{tabular}[c]{@{}l@{}}{\it Global} +{\it Local} (67 features)\\ \quad\quad\quad   IVG+IHVG\end{tabular}    & {\bf 83.5\%}    & {\bf 0.973}  & {\bf 83\%} \\ \hline
2D Hela descriptors                                                              &          & ${\bf N_{PCs}}$ & MAA   \\ \hline
\quad\quad\quad\quad P(k) IVG                                                              &          & 14      & 57.2\% \\
\quad\quad\quad\quad P(k) IHVG                                                             &          & 8      & 67.6\% \\
\quad\quad\quad\quad ${\bf Z}$ IVG                                                                  &          & 22      & 75.8\% \\
\quad\quad\quad\quad ${\bf Z}$ IHVG                                                               &          & 23      & 74\% \\ \hline
\end{tabular}
\caption{2D HeLa dataset. (Top) Best classification accuracy, best average by class AUC and model average classification accuracy MAA obtained with linear-kernel Multiclass one-vs-one Support Vector Machine (ovo-lSVM) classifier algorithm over 30 models realizations, for different sets of image visibility features. The best results are found when using a blend of global (degree distribution) and local features (visibility patches). (Bottom) The number of principal components ${\bf N_{PCs}}$ for each descriptor is tuned via exhaustive search in the space ${\bf N_{PCs}}\in[5,...,25]$ by maximizing the model average classification accuracy MMA over 30 model realizations.}
\label{tab:2}

\end{table}

\subsection{Image Visibility Multiplex for RGB Images: Natural Textures}

In natural scenes, more complicated textures can arise, e.g. patterns of waves on the surface of a lake or conformation and disposition of leaves from a plant when pictured from above. These kind of textures are generally classified as {\it{natural textures}} and are very relevant because they can be seen as the building blocks of more complex features arising in real world scenes.\\

\noindent Accordingly, in this subsection we consider three well known collections of natural textures: the Upstex dataset \cite{usptex}, the Colored Brodatz dataset \cite{brodatzbased}, and the Multiband (MTB) dataset \cite{brodatzbased}. The Upstex dataset includes 191 classes of 12 samples each showing textures from outdoor scenes such as walls, pavements, vegetation, soil, plant leaves, rocks (in Figure \ref{fig:datasets} c) we show some samples from different classes).
The Brodatz dataset in turn is among the most widely used datasets in the literature of texture analysis and comprehend scanned images from an album. In the colored version \cite{brodatzbased} of the dataset there are 112 images of size $640\times 640$, defining a total of 112 different classes (in Figure \ref{fig:datasets} c) we report some examples). For each class there are 16 samples defined as the non-overlapping subimages of size $160\times160$.    
Finally, the Multiband dataset is derived from the Brodatz dataset by aggregating triplets of original grayscale images in various combinations such that the resulting RGB images display a mixture of channel-dependent textures (see some examples in Figure \ref{fig:datasets} c)), totalling 154 classes and 16 samples per class. \\

\noindent All these three datasets are characterised by containing RGB images, many different classes, and a relatively small number of samples per class. Our approach here is to extract from each image the multiplex IVG/IHVG (by extracting the single-layer IVGs/IHVGs form each of the image channels: R,G,B) and obtain pseudo-multiplex descriptors by blending global/local features of each channel via PCA (see Figure \ref{fig:multiplex}). 
To obtain the highest classification performance we extracted global and local pseudo-multiplex features from both the multiplex IHVG and the multiplex IVG and we used a quadratic-kernel Multiclass one-vs-all Support Vector Machine (ova-qSVM), 10-fold cross validation.\\  
In order to optimize the number of components for each of the multiplex local/global features we follow a methodology analogous to the one used for the 2DHela dataset, but this time, to prove the efficiency of our classifier we tuned the number of principal components of the features only once for all the datasets by optimising the average classification accuracy of the model (MMA) over 30 model realizations measured on the sub-set of data of the Usptex dataset which includes only the first 50 classes (Figure \ref{fig:tuningPCs} (b)).
%he dimensionality of the space of global/local features in the multiplex scenario is higher and this justify t 
%The results confirm that global features extracted from IHVG are more informative than those from IVG ones, while for local features the opposite is true. 
%For a direct comparison of the performance of our approach with other approaches we train again our best model using the training set and the testing set suggested by the authors in 
In Table \ref{tab:3} we report for each of the datasets the values of accuracy, average AUC per class and MAA obtained with the model trained with global, local and global+local multiplex features.\\

\noindent Our best classification accuracy for Usptex, Brodatz and Multiband is respectively $98.4\%$, $99.8\%$, $99.6\%$ and is obtained with only 53 features that are a combination of multiplex local and global features from IHVG and IVG. Interestingly, in a very recent paper \cite{multilayerCN} the authors evaluated the performance of several pre-trained state-of-the-art convolutional neural network to two of these dataset (Usptex and Multiband), blended together with the performance obtained by extracting features from a  so-called Multilayer Complex Network (Multilayer CN) in a similar spirit with respect to our multiplex approach.  They report that $99.8\%$ as the best accuracy ever published for Usptex, that they obtain with a pre-trained ResNet50 2016, and  $97.1\%$ the best accuracy for Multiband obtained with their Multilayer Complex Network classifier. To the best of our knowledge we obtained the highest performance ever on the Multiband dataset, which suggests that our multiplex approach is ideal for detecting multi-band texture features and outperforms traditional CNN feedforward architectures, including the Multilayer CN proposed in \cite{multilayerCN}, that perhaps drop in performance because they are designed to learn texture features both from intra-band and inter-band pixel correlations. \\
Regarding the Brodatz colored dataset, we couldn't find any reference value of performance reported in literature, however for the grayscale original dataset a classification accuracy of $99.2\%$ was recently obtained via an hybrid classifier which combines a Convolutional Neural Network and a Support Vector Machine \cite{brodatzbest}.

\begin{table*}[]
\centering
\begin{tabular}{|l|l|l|l|}
\hline
\quad\quad\quad ova-qSVM                                                                       & Accuracy                                              & $\langle \text{AUC}\rangle$                                                  & MAA                                                   \\ \hline
\begin{tabular}[c]{@{}l@{}}{\it Global multiplex features:} P(k) in R,G,B + PCA\\ \quad\quad\quad IVG+IHVG (8 PCs+12 PCs)\\ Usptex\\ Brodatz Colored\\ Multiband\end{tabular} & \begin{tabular}[c]{@{}l@{}}\\ \\ 93.6\%\\ 98.7\%\\ 98.2\%\end{tabular} & \begin{tabular}[c]{@{}l@{}} \\ \\ 0.9829\\ 0.9929\\ 0.9982\end{tabular} & \begin{tabular}[c]{@{}l@{}} \\ \\93.2\%\\ 98.3\%\\ 97.8\%\end{tabular} \\ \hline

\begin{tabular}[c]{@{}l@{}}{\it Local multiplex features:} {\bf Z} in R,G,B + PCA\\ \quad\quad\quad IVG+IHVG (15 PCs+18 PCs)\\ Usptex\\ Brodatz Colored\\ Multiband\end{tabular} & \begin{tabular}[c]{@{}l@{}}\\ \\ 94.9\%\\ 98.6\%\\ 99.4\%\end{tabular} & \begin{tabular}[c]{@{}l@{}} \\ \\ 0.9952\\ 0.9967\\ 0.9994\end{tabular} & \begin{tabular}[c]{@{}l@{}} \\ \\94.3\%\\ 98.3\%\\ 99.2\%\end{tabular} \\ \hline

\begin{tabular}[c]{@{}l@{}}{\it Global} + {\it Local multiplex features} \\ \quad\quad\quad IVG+IHVG (53 features)\\ Usptex\\ Brodatz Colored\\ Multiband\end{tabular} & \begin{tabular}[c]{@{}l@{}}\\ \\ {\bf 98.4\%}\\ {\bf 99.8\%}\\ {\bf 99.6\%}\end{tabular} & \begin{tabular}[c]{@{}l@{}} \\ \\ 0.9975\\ 0.9993\\ 0.9996\end{tabular} & \begin{tabular}[c]{@{}l@{}} \\ \\98\%\\ 99.7\%\\ 99.4\%\end{tabular} \\ \hline

\end{tabular}
\caption{Natural Color Textures Datasets: best classification accuracy, best average by class AUC and model average classification accuracy MAA obtained with a quadratic-kernel Multiclass one-vs-all Support Vector Machine (ova-qSVM) classifier algorithm over 30 models realizations, for different sets of multiplex  image visibility features. Very slightly lower performance can be obtained by using a linear kernel SVM.}
\label{tab:3}
\end{table*}

\section{Conclusion}
The family of image visibility graphs IVG/IHVGs extract graphs from images and extend the concept of visibility graph time series analysis to image processing. In this paper we have showcased that this mapping can be leveraged to make (i) image filtering, (ii) compression and (iii) classification by extracting simple, universal (non-specific) and computationally efficient (i.e. fast and scalable extraction) features.\\ 

\noindent In a first step, we have shown how these graphs can be used to define filters for image preprocessing and image compression, which show good performance and are based on a methodology which is conceptually different from standard time/frequency domain filtering techniques.\\
Furthermore, by defining both global features (e.g. degree distribution) and local features (e.g. visibility patches, a novel tool which extends the concept of sequential visibility graph motifs \cite{motifs, motifs2} to images) we are able to reach competitive results in image classification tasks, sometimes superior to state of the art results obtained by sophisticated algorithms with highly specialised context-dependent descriptors. We found that global features seem to be more informative when extracted from the IHVG than from IVG, while the opposite is true for local features (e.g. patches), at least for low-order patches ($p=3$). We shall highlight that results are performed using low-order visibility patches ($p=3$), so better performance is likely to be obtained with higher order (e.g. $p=4$), for which linear time algorithms can also be designed \cite{motifs, motifs2}. An hybrid combination of our newly defined set of visibility graph features with more standard ad-hoc (i.e. problem-dependent) features and automatic feature extraction methods (e.g. artificial neural networks) is likely to provide even better results: this is an open problem for further research.\\ 

\noindent We should emphasize that the set of features extracted from IVG/IHVG --such as degree distributions or visibility patches-- is universal, in the sense that they are not designed to exploit particular, context-dependent properties of the image and can therefore be measured across classification tasks of different nature. Quoting Nanni \cite{nanni2}, ``The present trend in machine learning is focused on building optimal classification systems for very specific, well-defined problems. Another research focus, however, would be to work on building General-Purpose (GP) systems that are capable of handling a broader range of problems as well as multiple data types. Ideally, GP systems would work well out of the box, requiring little to no parameter tuning but would still perform competitively against less flexible systems that have been optimized for very specific problems and datasets.''\\
%\textcolor{red}{ COMM:To state the importance of general features see Nanni \cite{nanni2}: '' The present trend in machine learning is focused on building optimal classification systems for very specific, well-defined problems. Another research focus, however, would be to work on building General-Purpose (GP) systems that are capable of handling a broader range of problems as well as multiple data types. Ideally, GP systems would work well out of the box, requiring little to no parameter tuning but would still perform competitively against less flexible systems that have been optimized for very specific problems and datasets.''}
 Accordingly, we envisage that this methodology can have a potential impact in pattern recognition and image classification tasks across the disciplines, opening a plethora of further research directions. From a theoretical perspective, we recall that visibility graphs designed for time series analysis purposes have been shown previously to be amenable to analytic insight \cite{nonlinearity}. We conjecture that theoretical analysis --including analytical results-- for image visibility graphs is also possible.
%\textcolor{red}{ COMM: We obtain excellent results in texture classification and we think that the IVG approach can work in general for image classification tasks across the disciplines with more sophisticated architectures}
%textcolor{red}{ COMM: We have to stress that we get best highest accuracy some datasets) 
\noindent Other interesting open problems for further research include: (i) capitalizing on the properties of multiplex networks for extracting correlation-based multiplex features from RGB images, such as degree-degree correlations across layers \cite{battiston}, and the (ii) use the multiplex setting to study under a single framework videos (ordered sequence of images).\\

%\begin{itemize}
%\item IVGs and IHVGs can encapsulate structural information from images in a very robust way and extracting global and local features from those graphs to build up descriptors can be a very effective approach in machine learning applications such as pattern recognition and  image classification.
%\item IHVG can be used as filters for image pre-processing and image compression.
%\item Image visibility patches are universal local descriptors (exp. LBP depend on the image orientation) and do not depend on the nature of the dataset chosen. 
%\item The work opens up a pletora of further research directions such as exploring the potential of IVGs with state of the art AI classifiers like CNNs in real-world image classification challenges or studying non-square patches and multiplex patches, and extracting correlation-based multiplex features from RGB images such as degree-degree correlations across layers .
%\end{itemize}

\noindent {\bf Acknowledgments.} Thanks to N. Sadawi for interesting discussions. LL acknowledges funding from EPSRC Fellowship EP/P01660X/1. JI acknowledges funding from Wellcome Trust Metaboflow Grant 202952Glen.\\

%\begin{figure*}
%\centering
%\includegraphics[width=1.5\columnwidth]{image1}
%\caption{Example of an image and the same image after the visibility graph filter has been applied.}
%\label{image_example}
%\end{figure*}

%\noindent Once an I(H)VG has been extracted from a given image, one can extract standard topological properties of this graph. The degree $k$ of a node is the number of links of that node. The degree distribution $P(k)$ determines the probability of finding a node of degree $k$. The Shannon entropy over $P(k)$ $H=-\sum P(k)\log P(k)$. On the other hand, we can use any (local) topological feature, such that for a given $ij$, we can change ${\cal I}_{ij}$ by that property. This allows  to define a visibility filter:\\

%\noindent For illustrative purposes, in figure \ref{image_example} we show an example of a grey scale image and its associated horizontal visibility filter.

% SOME ML VENUES
% SIAM Review
% JMLR Journal of Machine learning research
% JAIR
% IEEE Transactions on Pattern Analysis and Machine Intelligence
% International Journal of Computer Vision

\bibliography{apssamp}% Produces the bibliography via BibTeX.

\end{document}